\documentclass[twocolumn,showpacs,nofootinbib,preprintnumbers,amsmath,amssymb,showkeys]{revtex4}

\usepackage{epsfig}
\usepackage{dsfont}
\usepackage{color}
\usepackage{graphicx}
\usepackage{dcolumn}
\usepackage{bm}

\newcommand{\QCD}{{\rm QCD}}
\newcommand{\FP}{{\rm FP}}
\newcommand{\tr}{{\rm tr\, }}

\newcommand{\sources}{{\rm src}}
\newcommand{\cb}{\bar c}
\newcommand{\Ab}{\bar A}
\newcommand{\Db}{\bar D}
\newcommand{\psib}{\bar \psi}

\def\slashii#1{\setbox0=\hbox{$#1$}             
   \dimen0=\wd0                                 
   \setbox1=\hbox{\sl/} \dimen1=\wd1            
   \ifdim\dimen0>\dimen1                        
      \rlap{\hbox to \dimen0{\hfil\sl/\hfil}}   
      #1                                        
   \else                                        
      \rlap{\hbox to \dimen1{\hfil$#1$\hfil}}   
      \hbox{\sl/}                               
   \fi}                                         %
\def\slashiii#1{\setbox0=\hbox{$#1$}#1\hskip-\wd0\hbox to\wd0{\hss\sl/\/\hss}}
\newcommand{\beq}{\begin{equation}}
\newcommand{\eeq}{\end{equation}}
\newcommand{\bea}{\begin{eqnarray}}
\newcommand{\eea}{\end{eqnarray}}
\newcommand{\nn}{\nonumber \\}

\newcommand\eqn[1]{(\ref{#1})}      
\newcommand\Eqn[1]{Eq.~(\ref{#1})}  
\newcommand\Fig[1]{Fig.~\ref{#1}}  
\newcommand\Sec[1]{Sec.~\ref{#1}}  

\begin{document}


\title{Perturbative study of the QCD phase diagram for heavy quarks\\ at nonzero chemical potential}

\author{U. Reinosa}%
\affiliation{%
Centre de Physique Th\'eorique, Ecole Polytechnique, CNRS, 91128 Palaiseau Cedex, France
}%
\author{J. Serreau}%
\affiliation{%
 Astro-Particule et Cosmologie (APC), CNRS UMR 7164, Universit\'e Paris 7 - Denis Diderot\\ 10, rue Alice Domon et L\'eonie Duquet, 75205 Paris Cedex 13, France
}%
\author{M. Tissier}
\affiliation{LPTMC, Laboratoire de Physique Th\'eorique de la Mati\`ere Condens\'ee, CNRS UMR 7600, Universit\'e Pierre et Marie Curie, \\ boite 121, 4 pl. Jussieu, 75252 Paris Cedex 05, France
}

\date{\today}

\begin{abstract}

We investigate the phase diagram of QCD with heavy quarks at finite temperature and chemical potential in the context of background field methods. In particular, we use a massive extension of the Landau-DeWitt gauge which is motivated by previous studies of the deconfinement phase transition in pure Yang-Mills theories. We show that a simple one-loop calculation is able to capture the richness of the phase diagram in the heavy quark region, both at real and imaginary chemical potential. Moreover, dimensionless ratios of quantities directly measurable in numerical simulations are in good agreement with lattice results.

 \end{abstract}

\pacs{12.38.Mh, 11.10.Wx, 12.38.Bx}
\keywords{QCD phase diagram, quantum field theory at finite temperature, deconfinement transition}
\maketitle


\section{Introduction}
\label{sec:intro}

The study of the phase diagram of QCD is an important theoretical challenge with numerous phenomenological implications, e.g.,  for heavy-ion collision experiments, early Universe cosmology, and astrophysics. The properties of the theory along the temperature axis have been intensively studied by means of lattice calculations and it is by now well established that the first-order confinement-deconfinement phase transition of the pure gauge SU($3$) theory becomes a crossover in the presence of light dynamical quarks with realistic masses \cite{Borsanyi:2013bia}. 

The situation is much less under control at finite baryonic chemical potential, where lattice simulations suffer from a severe sign problem \cite{deForcrand:2010ys,Philipsen:2010gj}. Calculations with realistic quark masses are limited to the region of small chemical potentials in units of the temperature. One hotly debated issue in this context is the possible existence of a line of first-order chiral transition with a critical endpoint, as predicted by various model calculations \cite{Stephanov:2004wx,deForcrand:2007rq,Schaefer:2007pw}. Efforts to tackle this question directly from the QCD action combine the investigation of various { ways to circumvent the sign problem on the lattice  \cite{deForcrand:2010ys,Sexty:2014zya,Aarts:2015kea,Tanizaki:2015pua} } and the use of nonperturbative continuum approaches \cite{Fischer:2011mz}, where the sign problem, although not completely absent, is much less severe. 

Unlike lattice calculations, continuum approaches rely on identifying efficient approximations for the relevant dynamics. Standard perturbative tools are notoriously inadequate in this context because the typical transition temperatures are of the order of the intrinsic scale of the theory, where the coupling becomes large. This motivates the use of nonperturbative approaches, typically based on truncations of the functional renormalization group or Dyson-Schwinger equations \cite{Fischer:2011mz,Fischer:2012vc,Fischer:2013eca,Herbst:2013ail,Gutierrez:2013sta,Fischer:2014ata,Fischer:2014vxa}. Lattice simulations in situations where they are well under control provide then valuable benchmarks for testing the various methods and approximations.

Examples of such situations are the case of a purely imaginary chemical potential, where there is no sign problem, or the study of QCD with heavy quarks. A systematic expansion around the infinite mass (quenched) limit can be devised, for which the sign problem is mild enough so that numerical simulations can be performed up to large chemical potentials \cite{de Forcrand:2002ci,de Forcrand:2003hx,D'Elia:2002gd,Fromm:2011qi}. Although these do not correspond to physically relevant situations, the phase diagram in these cases is interesting in its own and actually features a rich structure. {This has been investigated both on the lattice and by means of effective models \cite{Pisarski:2000eq,Dumitru:2003cf,Dumitru:2012fw,Kashiwa:2012wa,Kashiwa:2013rm,Nishimura:2014rxa,Nishimura:2014kla,Lo:2014vba}.}

In the present work, we investigate the phase diagram of QCD with heavy quarks by means of a modified perturbative expansion in the context of background field methods. 
This is based on a simple massive extension of the standard Faddeev-Popov (FP) action in the Landau-DeWitt gauge which has been successfully applied to the confinement-deconfinement transition of the SU($2$) and SU($3$) Yang-Mills theories \cite{Reinosa:2014ooa,Reinosa:2014zta,Reinosa:su3}. 

The motivations for such a massive extension are twofold. 
First, gauge-fixed lattice calculations of the vacuum Yang-Mills correlators in the (minimal) Landau gauge have shown that the gluon propagator behaves as that of a massive field at small momentum, whereas the ghost remains massless. This suggests that the dominant aspect of the nontrivial gluon dynamics (in the Landau gauge) might be efficiently captured by a simple mass term, up to corrections that can be computed perturbatively. This scenario has been put to test in Refs.~\cite{Tissier_10,Tissier_11}, where a one-loop calculation of the gluon and ghost propagators in a simple massive extension of the Landau gauge---known as the Curci-Ferrari model \cite{Curci76}---has been indeed shown to describe the lattice results remarkably well. Similar successful results have been obtained for the three-point correlation functions \cite{Pelaez:2013cpa} as well as for the vacuum propagators of QCD \cite{Pelaez:2014mxa} and the Yang-Mills propagators at finite temperature \cite{Reinosa:2013twa}. An interesting aspect of this approach is that the gluon mass acts as an infrared regulator and perturbation theory is well defined down to deep infrared momentum scales \cite{Tissier_11}.

The second motivation is more formal. It is related to the fact that the FP quantization procedure, which underlies standard perturbative tools, is plagued by the existence of Gribov ambiguities \cite{Gribov77} and is, at best, a valid description at high energies. For instance, it is known that a nonperturbative implementation of the BRST symmetry of the FP action on the lattice leads to undefined zero over zero ratios for gauge-invariant observables \cite{Neuberger:1986vv,Neuberger:1986xz}. In fact, existing gauge-fixing procedures on the lattice---e.g., the minimal Landau gauge---typically break the BRST symmetry. This suggests that a consistent quantization procedure, which correctly deals with the Gribov issue, is likely to induce effective BRST breaking terms. The simplest such term consistent with locality and renormalizability is a gluon mass term.\footnote{A more precise relation between the Gribov problem and the gluon mass term has been obtained in Ref.~\cite{Serreau:2012cg}, where a new quantization procedure based on a particular average over the Gribov copies along each gauge orbit has been proposed. The bare gluon mass is related to the gauge-fixing parameter which lifts the degeneracy between the copies.}

In Ref.~\cite{Reinosa:2014ooa}, we have extended this approach to the Landau-DeWitt gauge---which generalizes the Landau gauge in the presence of a nontrivial background field---with the aim of studying the confinement-deconfinement phase transition of SU($N$) Yang-Mills theories. Background field methods allow one to efficiently take into account the nontrivial order parameter of the transition~\cite{Braun:2007bx}. We have shown that a simple one-loop calculation correctly describes a confined phase at low temperature and a phase transition of second order for the SU($2$) theory and of first-order for the SU($3$) theory, with transition temperatures in qualitative agreement with known values. Two-loop corrections, computed in Refs.~\cite{Reinosa:2014zta,Reinosa:su3}, make this agreement more quantitative; see  Ref.~\cite{Serreau:2015saa} for a short review. { We mention that two-loop contributions also resolve some spurious unphysical behaviors (e.g., negative entropy) observed in background field calculations at one-loop order; see, e.g., \cite{Sasaki:2012bi}.}

The present work is a natural extension of these studies to QCD with $N_f=2+1$ heavy quarks flavors at finite chemical potential. We compute the background field potential at one-loop order, from which we can read the value of the order parameters---the averages of the traced Polyakov loop and of its Hermitic conjugate---as functions of the temperature and of the chemical potential. We show that this simple calculation accurately captures the rich structure of the phase diagram, both for real and imaginary chemical potential. Moreover, we obtain parameter-free values for the dimensionless ratios of the quark mass over the temperature at various critical points which compare well with lattice results.

The paper is organized as follows. We present the basic framework, i.e., the formulation of the QCD action in the massive extension of the Landau-DeWitt gauge in \Sec{sec:model} and discuss in detail various symmetry properties of the relevant generating function in \Sec{sec:sym}. This generalizes known material to the case of a nontrivial background field. The one-loop calculation of the background field potential is straightforward and is detailed in \Sec{sec:one_loop}. The rest of the paper is devoted to the  discussion of our results at vanishing (\Sec{sec:vanishing}),  imaginary (\Sec{sec:im}), and real (\Sec{sec:re}) chemical potential. In the latter case, we discuss how the sign problem also affects continuum approaches. Finally, Appendix~\ref{appeq:SSB} briefly presents some consequences of charge conjugation symmetry at vanishing chemical potential and Appendix~\ref{appsec:approx} provides an approximate calculation of the background field potential which allows for a simple analytic understanding of some results presented in the main text. 

\section{The QCD action in the massive Landau-DeWitt gauge}
\label{sec:model}

The Euclidean action of QCD in $d$ dimensions with $N$ colors and $N_f$ quark flavors reads 
\begin{equation}
\label{eq:action}
  S_\QCD=\int_x\bigg\{\frac 14 F_{\mu\nu}^aF_{\mu\nu}^a+\sum_{f=1}^{N_f}\psib_f( {\slashiii  {\cal D}} +M_f+\mu\gamma_0)\psi_f\bigg\},  
\end{equation}
where $\int_x=\int_0^\beta d\tau\int d^{d-1}x$, with $\beta$ the
inverse temperature, $F_{\mu\nu}^a=\partial_\mu A_\nu^a-\partial_\nu
A_\mu^a+g f^{abc}A_\mu^b A_\nu^c$, with $g$ the coupling constant and
$f^{abc}$ the structure constants of the SU($N$) group, and ${\cal D}_\mu\psi=\left(\partial_\mu-igA^a_\mu t^a\right)\psi$, with $t^a$ the SU($N$) generators in the fundamental representation, normalized as $\tr t^at^b=\delta^{ab}/2$. Finally, $\mu$ denotes the chemical potential. We leave the Dirac and color indices of the quark fields implicit and $\psi$ and $\psib$ are understood in the common sense as column and line bispinors, respectively. The Euclidean Dirac matrices\footnote{These are related to the standard Minkowski matrices as $\gamma_0=\gamma_M^0$ and $\gamma_i=-i\gamma_M^i$. In the following, we work in the Weyl  basis, where $\gamma_{0,2}^\star=\gamma_{0,2}^t=\gamma_{0,2}$ and $\gamma_{1,3}^\star=\gamma_{1,3}^t=-\gamma_{1,3}$.} $\gamma_\mu$ are Hermitian and satisfy the anticommutation relations
$\{\gamma_\mu,\gamma_\nu\}=2\delta_{\mu\nu}$. 

The gauge field $A_\mu^a$ is decomposed in a background field $\Ab_\mu^a$ and a
fluctuating contribution as
\begin{equation}
  \label{eq:Aba}
  A_\mu^a=\Ab_\mu^a+a_\mu^a\,,
\end{equation}
and the Landau-DeWitt gauge is defined as
\beq
  (\Db_\mu a_\mu)^a=0,
 \eeq
 with $\Db^{ab}_\mu=\delta^{ab}\partial_\mu+ gf^{acb}\Ab_\mu^c$ the background covariant derivative in the adjoint representation. The corresponding Faddeev-Popov gauge-fixing action reads
\begin{equation}
  \label{eq:fp}
  \begin{split}
    S_\FP=\int_x \Big\{(\Db_\mu \cb)^a(D_\mu c)^a+ih^a(\Db_\mu a_\mu)^a\Big\}\,,
  \end{split}
\end{equation}
where $D^{ab}=\delta^{ab}\partial_\mu+ gf^{acb}A_\mu^c$, $c$ and $\cb$ are anticommuting ghost fields,
and $h$ is a Lagrange multiplier. Finally, as discussed in the Introduction, we also consider a massive extension of the Landau-DeWitt gauge, with mass term\footnote{ In principle, one could implement different bare masses for the gluon fields in the temporal and spatial directions, e.g., to account for Gribov ambiguities on asymmetric lattices at finite temperature. We chose here the minimal extension of the FP theory with a symmetric, vacuumlike mass term.}~\cite{Reinosa:2014ooa}
\begin{equation}
  \label{eq:mass}
  S_m=\int_x \frac 12 m^2 a_\mu^a a_\mu^a\,.
\end{equation}
 In the following, we grab together the fluctuating fields of the pure gauge sector, which all belong to the adjoint representation of the gauge group, as $\varphi\equiv(a_\mu,c,\cb,h)$. The total gauge-fixed action $S_\QCD+S_\FP+S_m$ is invariant under the combined transformation of the fluctuating fields 
\beq
\label{eq:local1}
 \psi\to U\psi\,,\quad\psib\to\psib U^\dagger\,,\quad\varphi^at^a\to U\varphi^at^a U^\dagger
\eeq
and of the background field
\beq
\label{eq:local2}
 \Ab_\mu^at^a\to U\Ab_\mu^at^a U^\dagger+\frac{i}{g}U\partial_\mu U^\dagger,
\eeq
where $U$ is a local SU($N$) matrix.

The background field $\Ab$ is nothing but a gauge-fixing parameter and is used here as a device to capture nontrivial physics in an efficient way. We constrain it so as to break as few symmetries as possible. The symmetries of the Euclidean space with the boundary conditions implied by finite temperature field theory allow for a constant vector field in the temporal direction: $\Ab_\mu^a(x)=\delta_{\mu0}\bar A^a$. Moreover, it is always possible, through a global color transformation, to bring the color vector $\Ab^a$ in the Cartan subalgebra of the gauge group, spanned by the diagonal generators. In the following, we denote the latter by $t^k$. For SU($3$), these are $t^3=\lambda^3/2$ and $t^8=\lambda^8/2$, where $\lambda^a$ are the Gell-Mann matrices. Accordingly, we  consider the generating function (we make explicit the dependence on the chemical potential for latter use)
\begin{equation}
  \label{eq_Z}
  Z(J,\Ab,\mu)=\int\mathcal D(\varphi,\psi,\bar\psi)e^{-S_\QCD-S_\FP-S_m+S_\sources}\,,
\end{equation}
with the source term
\begin{equation}
  \label{eq:source}
  S_\sources=\int_x J_\mu^a A_\mu^a
\end{equation}
restricted to a constant source in the temporal direction and in the Cartan subalgebra: $J_\mu^a(x)=\delta_{\mu0}J^a$, with $a=3,8$. In the following, $J$ and $\Ab$ denote vectors in the plane spanned by the SU($3$) Cartan directions (to which we shall refer as the Cartan plane in the following):
\beq
 J=\left(\begin{tabular}{c}$J^3$\\$J^8$\end{tabular}\right)\,,\quad \Ab=\left(\begin{tabular}{c}$\Ab^3$\\$\Ab^8$\end{tabular}\right).
\eeq
The main quantity of interest in the following is the Legendre transform $-\ln Z\left(J(A,\bar A,\mu),\Ab,\mu\right)+\int_x J^k(A,\Ab,\mu)A^k$, where $J^k(A,\Ab,\mu)$ is defined through 
\beq
 \label{eq:sourceAJA}
 A^k=\frac{1}{\beta\Omega}\left.\frac{\partial\ln Z}{\partial J^k}\right|_{J(A,\bar A,\mu)}\,.
\eeq
We shall evaluate this Legendre transform at $A=\bar A$, that is,
\begin{equation}
  \label{eq_pot}
  V(r,\mu)=-\frac{1}{\beta\Omega}\ln Z\!\left(\tilde J(\bar A,\mu),\Ab,\mu\right)+\tilde J^k(\Ab,\mu)\Ab^k\,,
\end{equation}
where $r^k=\Ab^k/(\beta g)$, $\Omega$ is the spatial volume, and where we defined $\tilde J^k(\Ab,\mu)=J^k(\Ab,\Ab,\mu)$. {\it A priori}, one could leave $A$ independent of $\bar A$ and extremize the effective potential with respect to $A$ at fixed $\Ab$. However, one obtains the same physics by identifying $A=\Ab$ and extremizing\footnote{We shall see below that, for any imaginary chemical potential, one needs to minimize the potential \eqn{eq_pot}. However, for real chemical potential, some amendments need to be made to this rule.} with respect to $\Ab$. This is a particularly convenient choice since the fluctuating gauge field has vanishing expectation value, $\langle a\rangle=0$, even at nonvanishing source.

Finally, important physical observables to be considered below are the averages of the traced Polyakov loop in the fundamental representation and of its Hermitic conjugate:
\begin{align}
\label{eq_popoldef}
\ell(\mu)&=\frac 13\tr \left\langle P \exp\left(i g\int_0^\beta \!d\tau A_0^at^a\right)\right\rangle\,,\\
\label{eq_popoldefbar}
\bar \ell(\mu)&=\frac 13\tr \left\langle \bar P \exp\left(-i g\int_0^\beta \!d\tau A_0^at^a\right) \right\rangle\,,
\end{align}
where $P$ and $\bar P$ denote path ordering and antiordering respectively. In general, $\bar \ell(\mu)$ is not equal to the complex conjugate $\ell^\star(\mu)$, e.g., in the case of a complex action. For real chemical potential, the quantities \eqn{eq_popoldef} and \eqn{eq_popoldefbar} are real and are  related to the free energies $F_q$ and $F_{\bar q}$ of a static quark or antiquark as $\ell(\mu)\propto \exp(-\beta F_q)$ and $\bar\ell(\mu)\propto \exp(-\beta F_{\bar q})$ \cite{Svetitsky:1985ye,Philipsen:2010gj}. 
The physical loops $\ell(\mu)$ and $\bar\ell(\mu)$ are to be evaluated at vanishing sources, that is, at an extremum of the background field potential \eqn{eq_pot}. In order to discuss symmetries below, it is useful to introduce the following functions of the background field, defined at nonvanishing source $\tilde J(\Ab,\mu)\neq0$,
\begin{align}
\label{eq_popol}
\ell(r,\mu)&=\frac 13 \tr\left\langle P \exp\left(ir^kt^k+ig\int_0^\beta \!\!d\tau \,a_0^at^a\right)\right\rangle\,,\\
\label{eq_popolbar}
\bar\ell(r,\mu)&=\frac 13 \tr\left\langle\bar  P \exp\left(-ir^kt^k-ig\int_0^\beta \!\!d\tau\, a_0^at^a\right)\right\rangle,  
\end{align}
where it is understood that the averages on the right-hand side are evaluated at $A=\Ab$, that is, at $\langle a\rangle=0$. The physical loops $\ell(\mu)$ and $\bar\ell(\mu)$ are obtained by evaluating the functions \eqn{eq_popol} and \eqn{eq_popolbar} at the relevant extremum of the potential \eqn{eq_pot}.

\section{Symmetries}
\label{sec:sym}

In this section, we discuss various transformation properties of the generating function \eqn{eq_Z}, background field potential \eqn{eq_pot} and Polyakov loops functions \eqn{eq_popol} and \eqn{eq_popolbar}. For later purposes, it is useful to consider the general case where $J^k$, $\Ab^k$, and $\mu$ are complex numbers. We first discuss the effect of charge conjugation and of complex conjugation which relate the cases $\mu\leftrightarrow-\mu$ and $\mu\leftrightarrow\mu^\star$, respectively. Next, we analyze the consequences of the continuous global and local color transformations. Finally, we discuss the Roberge-Weiss symmetry which relates the theories with chemical potentials $\mu$ and $\mu+2i\pi/(3\beta)$.

\subsection{Charge conjugation}
\label{sec_muto-mu}
We perform the change of variables
\beq
\label{eq:C}
  \psi\to {\cal C}\bar\psi^t\,,\quad\bar\psi\to -\psi^t{\cal C}^\dagger\,,\quad\varphi\to C\varphi
\eeq
under the functional integral \eqn{eq_Z}. Here the charge conjugation matrix ${\cal C}$ only acts on the Dirac indices of fields in the fundamental representation whereas $C$ acts on the color indices of adjoint fields. The former satisfies 
\beq
 {\cal C}^\dagger\gamma_\mu{\cal C}=-\gamma_\mu^t
\eeq
and is given by ${\cal C}=\gamma_0\gamma_{2}$ in the Weyl basis. It is such that ${\cal C}^{-1}={\cal C}^\dagger={\cal C}^t=-{\cal C}$. As for the latter, it acts on the generators of the color group as
\beq
 C^{ab}t^b=-\left(t^a\right)^t.
\eeq
Using the Gell-Mann basis, one easily checks that it is a diagonal matrix with eigenvalues $+1$ on the lines 2, 5 and 7 and $-1$
otherwise. It is thus an orthogonal $O(8)$ matrix such that ${C}^{-1}={C}^t={C}$ and which conserves the
structure constants ($C^{aa'}C^{bb'}C^{cc'}f^{a'b'c'}=f^{abc}$). Notice, however, that  it has
determinant $-1$ and it is, therefore, not a color transformation in the
adjoint representation. 

It is an easy exercise to check that the only effect of the change of variables \eqn{eq:C} on the generating function \eqn{eq_Z} is to change
\beq
 J\to CJ=-J\,,\quad\Ab\to C\Ab=-\Ab\,,\quad{\rm and}\quad\mu\to-\mu,
\eeq
where we used the fact that $C^{kk'}=-\delta^{kk'}$ in the Cartan subalgebra.  We conclude that
\begin{equation}
  \label{eq_muto-mu}
Z(J,\Ab,\mu)=Z(-J,-\Ab,-\mu)\,. 
\end{equation}
This implies the relation $J(-A,- \bar A,-\mu)=-J(A,\bar A,\mu)$ for the source defined in \Eqn{eq:sourceAJA}. It follows that $\tilde J(- \bar A,-\mu)=-\tilde J(\bar A,\mu)$ and thus
\beq
\label{eq:VC}
 V(r,\mu)=V(-r,-\mu).
\eeq
Similarly, we deduce the relation
\begin{equation}
\label{eq_polmuto-mu}
  \ell(r,\mu)=\bar\ell(-r,-\mu).
\end{equation}

\subsection{Complex conjugation}
\label{sec_complex}

We now consider the change of variables\footnote{This is similar to the ${\cal K}$-transformation discussed in Refs.~\cite{Nishimura:2014rxa,Nishimura:2014kla}.}  
\begin{align}
  \psi\to {\cal K}\psi\,,\quad\bar\psi\to \psib {\cal K}^\dagger\,,\quad\varphi\to K\varphi,
\end{align}
where the matrix ${\cal K}=\gamma_1\gamma_3$ only acts on Dirac indices and is such that ${\cal K}^{-1}={\cal K}^\dagger={\cal K}^t=-{\cal K}$ and
\beq
 {\cal K}^\dagger\gamma_\mu{\cal K}=\gamma_\mu^\star.
\eeq
As for  the matrix $K$ acting on the color indices of the adjoint fields, it satisfies
\beq
 K^{ab}t^b=-\left(t^a\right)^\star.
\eeq
Using the fact that the generators $t^a$ are Hermitian, we see that $K=C$.
Upon the additional change of integration variable $h^a\to-h^a$ and the transformation
\beq
  J\to KJ^\star=-J^\star\,,\quad\Ab\to K\Ab^\star=-\Ab^\star\,,\quad{\rm and}\quad\mu\to\mu^\star,
\eeq
one recovers the original form of the generating function with all complex coefficients replaced by their complex conjugate, that is,\footnote{Note that the integration variables are not affected by the complex conjugation.}
\begin{equation}
  \label{eq_Zstar}
  Z\left(J,\Ab,\mu\right)=Z^\star\!\left(-J^\star,-\Ab^\star,\mu^\star\right).
\end{equation}
Following similar steps as before, this implies $\tilde J(- \bar A^\star,\mu^\star)=-\tilde J^\star(\bar A,\mu)$ and thus
\beq
\label{eq:VK}
 V(r,\mu)=V^\star(-r^\star,\mu^\star).
\eeq
One also shows that
\beq
\label{eq_polconj}
  \ell(r,\mu)=\ell^\star(-r^\star,\mu^\star)\\
\eeq
and similarly for $\bar\ell$.

\subsection{Global color symmetry}
\label{sec_color} 

The gauge-fixed action $S_\QCD+S_\FP+S_m$ is invariant under global SU($3$) transformations 
\beq
 \psi\to {\cal R}\psi\,,\quad\psib\to\psib {\cal R}^\dagger\,,\quad\varphi\to R\varphi,
\eeq
 where ${\cal R}$ is a global SU($3$) matrix and $R$ the corresponding SO(8) transformation 
\begin{equation}
  R^{ab}=2\, \tr \left({\cal R}^\dagger t^a {\cal R} t^b\right)\,.
\end{equation}
Of particular interest are those global color transformations which do not mix the Cartan subalgebra with the other elements of the Lie algebra (i.e., $R$ is block-diagonal), such that they leave the source $J$ and the background $\Ab$ in the Cartan subalgebra. For instance, the SU($3$) matrix
\begin{equation}
  \label{eq_s1}
  {\cal R}_1=
  \begin{pmatrix}
    $0$ & $1$ & $0$\\
   $-1$ & $0$ & $0$\\
    $0$ & $0$ & $1$
  \end{pmatrix}
\end{equation}
induces such a block-diagonal color rotation $R_1$ whose restriction to the Cartan subalgebra reads
\beq
\label{eq_R1}
 R_1^{\rm Cartan}={\rm diag}(-1,1).
\eeq
In the Cartan plane, this corresponds to a reflexion about the $8$-axis. 
We can therefore write
\begin{equation}
  \label{eq_reflexion}
  Z\left(J, \Ab,\mu\right)=Z\left(R_1J, R_1\Ab,\mu\right)\,.
\end{equation}

As before, this implies that the source $\tilde J(\Ab,\mu)$ transforms covariantly, i.e., $\tilde J(R_1\bar A,\mu)=R_1\tilde J(\bar A,\mu)$, from which we conclude that
\beq
\label{eq_Vreflexion}
 V(r,\mu)=V(R_1 r,\mu).
\eeq
Similar considerations yield, for the functions \eqn{eq_popol} and \eqn{eq_popolbar},
\beq
\label{eq_polreflexion}
\ell(r,\mu)=\ell(R_1r,\mu)
\eeq
and similarly for $\bar\ell$.

It is easy to check  that the SU($3$) matrices
\begin{equation}
  \label{eq_mat_mirror2}
    {\cal R}_2=
  \begin{pmatrix}
    $0$ & $0$ & $-1$\\
    $0$ & $1$ & $0$\\
    $1$ & $0$ & $0$
  \end{pmatrix}\quad\text{and}\quad
    {\cal R}_3=
  \begin{pmatrix}
    $1$ & $0$ & $0$\\
    $0$ & $0$ & $-1$\\
    $0$ & $1$ &$0$
  \end{pmatrix}
\end{equation}
also generate color rotations in the adjoint representation such that the Cartan components do not mix with other color directions. In the Cartan plane, they correspond to mirror images about two axes passing through the origin and making an angle of $\pm \pi/6$ with the $3$-axis. Together with $R_1$, they generate all the color rotations under which the Cartan subalgebra is stable. These form a $C_{3v}$ group which also contains rotations by an angle $\pm 2\pi/3$ around the origin. This is the Weyl group of the su($3$) algebra.

\subsection{Background gauge invariance}
\label{sec_perio}

We now come to analyze the consequences of the invariance of the action under the local transformation \eqn{eq:local1}--\eqn{eq:local2}. In order to maintain the background field  constant, in the temporal direction, and in the Cartan subalgebra, we consider local color transformations of the form $U(\tau)=e^{i(\tau/\beta) \phi^kt^k}$, which simply generate a translation of the background field in the Cartan plane: $\Ab\to\Ab+\phi/(\beta g)$. True gauge transformations---which leave physical observables invariant---are periodic in time, 
$U(\beta)=U(0)$, which implies that the eigenvalues of the matrix $\phi^kt^k$ must be multiples of $2\pi$. This is solved by
\beq
\label{eq:trans}
 \phi=j_1 u_1+j_2 u_2,
\eeq
with $j_{1,2}$ integers and 
\begin{equation}
  \label{eq_vecs}
  u_1=2\pi
  \begin{pmatrix}
     1\\
      {\sqrt 3}
  \end{pmatrix}\,,\quad
  u_2=2\pi
  \begin{pmatrix}
     1\\
      {-\sqrt 3}
  \end{pmatrix}.
\end{equation}
Both the integration measure and the action at vanishing source in \eqn{eq_Z} are left invariant by the transformation \eqn{eq:local1}-\eqn{eq:local2}. The change of the source term is trivial and we get
\beq
\label{eq_shiftsource}
 \frac{\ln Z\!\left(J,\bar A+\phi/(\beta g),\mu\right)}{\beta\Omega}=\frac{\ln Z(J,\bar A,\mu)}{\beta\Omega}+\frac{J^k\phi^k}{\beta g}.
\eeq
We deduce that $\tilde J\left(\bar A+\phi/(\beta g),\mu\right)=\tilde J(\bar A,\mu)$ and thus that
\begin{equation}
  \label{eq_translat}
  V(r,\mu)=V( r+\phi,\mu)\,,
\end{equation}
with $\phi$ given in \Eqn{eq:trans}. One also shows that
\begin{equation}
\label{eq_poltrans}
  \ell(r,\mu)=\ell(r+\phi,\mu)
\end{equation}
and similarly for $\bar\ell$.

\subsection{Roberge-Weiss symmetry}
\label{sec_center}
Following Ref.~\cite{Roberge:1986mm}, we now consider local SU($3$) transformations of the type studied in the
previous subsection with $U(\tau)$ periodic in time up to a nontrivial element of the center $Z_3$ of the gauge group: $U(\beta)=e^{ \pm2i\pi/3}U(0)$. This results in translations $\Ab\to\Ab+j_1 e_1+j_2 e_2$ of the background field with $j_{1,2}$ integers and\footnote{The vectors $e_1$ and $e_2$ generate a lattice dual to that generated by the roots of the algebra; see below. This property generalizes to any compact Lie group with a simple Lie algebra, see \cite{Reinosa:su3} for a detailed discussion.}
\beq
   \label{eq_vecs2}
  e_1={2\pi}
  \begin{pmatrix}
     1\\
     - 1/ {\sqrt 3}
  \end{pmatrix},
\qquad   e_2={2\pi}
  \begin{pmatrix}
    -1\\
    -1/  {\sqrt 3}
  \end{pmatrix}.
\eeq
 As is well known, the pure gauge action---including the FP and mass terms---is invariant under such twisted gauge transformations, whereas the source term transforms trivially, as in \Eqn{eq_shiftsource}. This symmetry is explicitly broken by the antiperiodic boundary conditions of the quark fields in the fundamental representation. To cope for this, we perform the additional change of variables
\begin{equation}
  \psi\to e^{\mp\frac{2i\pi}3\frac{\tau}\beta}\psi\,,\quad\psib\to e^{\pm\frac{2i\pi}3\frac{\tau}\beta}\psib.
\end{equation}
All terms in the functional integral \eqn{eq_Z} are invariant except for the quark kinetic term, whose variation, $\mp\frac{2i\pi}{3\beta}\psib\gamma_0\psi$, can, however, be compensated by a shift of the chemical potential in the imaginary direction. A similar analysis as in the previous subsections leads to the following identities for the background field potential
\begin{align}
\label{eq_center}
V(r,\mu)&=V\left(r+ e_j,\mu+\frac{2i\pi}{3\beta}\right).
\end{align}
where $j=1,2$. As is well known, the traced Polyakov loop gets multiplied by an element of the center under the twisted gauge transformations considered here. This leads to
\begin{align}
\label{eq_polcenter}
  \ell\left(r,\mu\right)&= e^{-2i\pi/3}\ell\left(r+e_{j},\mu+\frac{2 i\pi}{3\beta}\right),\\
  \bar\ell\left(r,\mu\right)&= e^{ 2i\pi/3}\bar\ell\left(r+e_{j},\mu+\frac{2 i\pi}{3\beta}\right).
\end{align}

Observe that $e_1-e_2=u_1+u_2$, which shows that if we use the property \eqn{eq_center} twice so as to add and subtract $2i\pi/(3\beta)$ to the chemical potential, we retrieve the original potential, up to a gauge transformation of the form \eqn{eq_translat}. Similarly, using the property \eqn{eq_center} thrice so as to add  three times $2i\pi/(3\beta)$ to the chemical potential and using the relation $2e_1+e_2=u_2$ we see that we generate a gauge transformation of the form \eqn{eq_translat} up to a translation by $2i\pi/\beta$ of the chemical potential. We, thus, recover the well-known fact that the physics is unaffected by a change\footnote{This can be checked by performing the change of variables $\psi\to e^{2i\pi\tau/\beta}\psi$, $\bar\psi\to e^{-2i\pi\tau/\beta}\bar\psi$, which preserves the antiperiodic boundary conditions.} $\mu\to\mu+2i\pi/\beta$.

\subsection{Vanishing sources}

We end this section by discussing the case of vanishing sources $J=0$, which corresponds to the physical point. For instance, the partition function $Z(\mu)$ is obtained as the generating function \eqn{eq_Z} at $J=0$. In the present context, where we set $A=\Ab$, it is obtained from \Eqn{eq_pot} at $\tilde J=0$, namely,
\beq
\label{eq:partitionfunction}
 -\frac{\ln Z(\mu)}{\beta\Omega}=V\big(r_{\rm ext}(\mu),\mu\big)
\eeq
where $r_{\rm ext}(\mu)$ is any physical extremum of $V(r,\mu)$. In general, there are many such extrema, which form a set $\{r_{\rm ext}(\mu)\}$. The properties \eqn{eq:VC} and \eqn{eq_center} imply\footnote{This is obviously true for the set of absolute minima which are the relevant extrema at vanishing and imaginary chemical potential. For a real chemical potential, the choice of the relevant extrema is more subtle, as discussed below. Our prescription, in Sec.~\ref{sec:re}, is also compatible with \Eqn{eq:partitionfunc}.} that this set satisfies, 
\beq
 \{r_{\rm ext}(\mu)\}=\{-r_{\rm ext}(-\mu)\}=\left\{r_{\rm ext}\big(\mu\pm2i\pi/(3\beta)\big)\mp e_j\right\},
\eeq
with $j=1,2$.
Similarly, the relation \eqn{eq:VK} implies that the set of extrema of $V^\star(r,\mu^\star)$ is $\{-r^\star_{\rm ext}(\mu^\star)\}$. Using \Eqn{eq:partitionfunction} and the relations \eqn{eq:VC}, \eqn{eq:VK}, and \eqn{eq_center}, we then retrieve the standard relations for the partition function \cite{Philipsen:2010gj} 
\beq
\label{eq:partitionfunc}
 Z(\mu)=Z(-\mu)=Z^\star(\mu^\star)=Z\left(\mu+\frac{2i\pi}{3\beta}\right).
\eeq
Similarly, using the definitions $\ell(\mu)=\ell\big(r_{\rm ext}(\mu),\mu\big)$ and $\bar \ell(\mu)=\bar\ell\big(r_{\rm ext}(\mu),\mu\big)$ and assuming that none of the symmetries mentioned here is spontaneously broken, we get the known relations for the averages of the traced Polyakov loop and of its Hermitic conjugate:
\beq
\label{eq:ellsym}
 \ell(\mu)=\bar\ell(-\mu)=\ell^\star(\mu^\star)=e^{-2i\pi/3}\ell\left(\mu+\frac{2i\pi}{3\beta}\right).
\eeq
For imaginary chemical potential, the Roberge-Weiss symmetry can be spontaneously broken. In that case, the averaged Polyakov loops can be multivalued and the relations in \eqn{eq:ellsym} relate the sets of their possible values.

\section{The background field potential at one-loop order}
\label{sec:one_loop}

Here, we detail the calculation of the leading-order, one-loop effective potential \eqn{eq_pot}. We consider a general compact Lie group with simple Lie algebra and we specify to SU($3$) when needed. For completeness we also give the corresponding leading-order (tree-level) expressions of the Polyakov loop functions \eqn{eq_popol} and \eqn{eq_popolbar}. Our approach is similar to that used in \cite{Dumitru:2012fw}.

The one-loop contribution to the potential only involves the action up to quadratic order in the fluctuating fields. The pure gauge contribution has been computed in Ref.~\cite{Reinosa:2014ooa}. Denoting by $M$ the quark mass matrix, the relevant contribution from the quark sector is 
\beq
\label{eq:1loopq}
 V_q(r,\mu)=-{\rm Tr}\,{\rm Ln}\left(\slashiii  \partial +M+\mu\gamma_0-i g \gamma_0\Ab^k t^k\right),
\eeq
where the functional trace ${\rm Tr}$ involves a sum over fermionic Matsubara frequencies, an integral over spatial momenta and a trace over Dirac, color, and flavor indices. The Cartan generators $t^k$ can be diagonalized simultaneously and their respective eigenvalues $\rho^k$ form a set of $d_F$ (possibly degenerate, i.e., identical) vectors $\rho$ in the space spanned by the Cartan directions, with $d_F$ the dimension of the fundamental representation.\footnote{The present calculation easily generalizes to fields in any representation, provided one works with the corresponding weights.} In group theory language, these are called the weights of the representation. The fundamental representation of the su($3$) algebra has three nondegenerate weights, $\rho\in\{\rho_1,\rho_2,\rho_3\}$, which are related to the vectors $e_{1,2}$ introduced in \Eqn{eq_vecs2} as $\rho_{1,2}=-e_{1,2}/(4\pi)$ and $\rho_1+\rho_2+\rho_3=0$. Explicitly,
\begin{align}
\rho_1&=\frac{1}{2}\left(\begin{array}{c}
-1\\
1/\sqrt{3}
\end{array}\right),\\ 
\rho_2&=\frac{1}{2}\left(\begin{array}{c}
1\\
1/\sqrt{3}
\end{array}\right),\\
\rho_3&=-\left(\begin{array}{c}
0\\
1/\sqrt{3}
\end{array}\right)\,.
\end{align}
Having diagonalized the color structure in \Eqn{eq:1loopq}, we see that the constant temporal background $\Ab$ can be absorbed in a redefinition of the chemical potential $\mu\to\mu-ir_\rho/\beta$, with $r_\rho=r^k\rho^k$, which lifts the degeneracy between the various color states. We thus get
\begin{equation}
  \label{eq_potq1loop}
  V_q(r,\mu)=\sum_{f,\rho}V_f^0(\mu-ir_\rho/\beta),
\end{equation}
where the sum runs over all flavors and all color states (weights $\rho$) in the fundamental representation and $V_f^0(\mu)$ is the one-loop contribution from a single quark flavor $f$ in a definite color state at vanishing background field and nonzero chemical potential $\mu$ {  \cite{Kashiwa:2012wa}}. This is given by the standard formula (written here for $d=4$  and letting aside a trivial $T$- and $\mu$-independent piece)
\beq
\label{eq:V0f}
V_{f}^0(\mu) \!=\! -\frac{T}{\pi^2}\!\int_0^\infty \!\!dq\,q^2\!\left\{\ln\!\left[1+e^{-\beta\left(\varepsilon_q^f+\mu\right)}\right]\!+\!(\mu\to-\mu)\right\}\!,
\eeq
where $\varepsilon_q^f=\sqrt{q^2+M_f^2}$.

The pure gauge contribution has been computed in \cite{Reinosa:2014ooa} for SU($2$) and SU($3$). It is instructive to derive the corresponding formula for a general  compact Lie group with simple Lie algebra. After taking into account the contribution from the massive gluon modes and the partial cancelation between massless modes contributions from the ghost and the $h-a$ sectors, one has, in $d=4$,
\beq
\label{eq:onelooppot}
 V_{\rm gauge}(r)=\frac{3}{2}{\cal F}_m(r)-\frac{1}{2}{\cal F}_0(r),
\eeq
where
\beq
 {\cal F}_m(r)={\rm Tr}\,{\rm Ln}\left(\bar D^2+m^2\right)
\eeq
and $\bar D_\mu=\partial_\mu-ig\delta_{\mu0}\bar A^k T^k$, with $T^k$ the generators of the Lie algebra in the adjoint representation. Here, the functional trace ${\rm Tr}$ involves a sum over bosonic Matsubara frequencies, a spatial integral and a trace over color indices.

As for the quark contribution, this trace can be easily evaluated using the weights of the adjoint representation. In the subspace spanned by the Cartan directions, these form a set of $d_A$ vectors $\kappa$ whose components are the eigenvalues of the Cartan generators $T^k$, where $d_A$ is the dimension of the adjoint representation. One sees that the temporal background  enters as a shift of the Matsubara frequencies, which can be interpreted as an effective imaginary chemical potential $-ir_\kappa/\beta$, with $r_\kappa=r^k\kappa^k$. We thus get
\begin{align}
\label{eq:Weiss00}
 {\cal F}_m(r)=\sum_{\kappa}\frac{T}{\pi^2}\int_0^\infty \!\!dq\,q^2\ln\left[1-e^{-\beta\varepsilon_q+ir_\kappa}\right],
\end{align}
where the sum runs over all color states (weights $\kappa$) in the adjoint representation. By definition, the generators $T^k$ have as many zero eigenvalues---corresponding to $\kappa=0$---as there are Cartan directions. Furthermore, one can show that the nonzero weights always come in pairs $\kappa=\pm\alpha$, which are called the roots of the Lie algebra. We obtain,
\begin{align}
\label{eq:Weiss0}
 {\cal F}_m(r)=\frac{T}{\pi^2}\!&\int_0^\infty \!\!dq\,q^2\Big\{d_C \ln\left[1-e^{-\beta\varepsilon_q}\right]\nn
 +&\sum_{\alpha}\ln\left[1+e^{-2\beta\varepsilon_q}-2e^{-\beta\varepsilon_q}\cos r_\alpha\right]\!\Big\},
\end{align}
where $d_C$ is the dimension of the Cartan subalgebra and the sum runs over the $(d_A-d_C)/2$ possible pairs $(\alpha,-\alpha)$. For SU($3$), there are two Cartan directions and $d_A=8$ so there are six roots $\alpha\in\{\pm\alpha_1,\pm\alpha_2,\pm\alpha_3\}$, where we can choose [see \Eqn{eq_vecs}] $\alpha_{1,2}=u_{1,2}/(4\pi)$ and $\alpha_1+\alpha_2+\alpha_3=0$. Explicitly,
\begin{align}
\alpha_1&= \frac{1}{2}\left(\begin{array}{c}
1\\
\sqrt{3}
\end{array}\right),\\
\alpha_2&= \frac{1}{2}\left(\begin{array}{c}
1\\
-\sqrt{3}
\end{array}\right),\\
\alpha_3&=-\left(\begin{array}{c}
1\\
0
\end{array}\right).
\end{align}
The sum in \eqn{eq:Weiss0} runs only over the set $\{\alpha_1,\alpha_2,\alpha_3\}$. For completeness, we mention that, for $r_\alpha\in[0,2\pi]$, the function ${\cal F}_{m=0}(r)$ admits the closed form \cite{Weiss:1980rj,Gross:1980br}
\begin{align}
\label{eq:Weiss}
{\cal F}_0(r) & = -\frac{\pi^2T^4}{45}d_C\nonumber\\
& + \frac{T^4}{6}\sum_{\alpha}\left[\frac{(r_\alpha-\pi)^4}{2\pi^2}-(r_\alpha-\pi)^2+\frac{7\pi^2}{30}\right].
\end{align}

Finally, the leading-order, tree-level expressions of the Polyakov loop functions \eqn{eq_popol} and \eqn{eq_popolbar} are trivially obtained as 
\beq
\label{eq:popolsu3general}
 \ell(r,\mu)=\bar\ell(-r,\mu)=\frac{\tr \big(e^{ir^kt^k}\big)}{d_F}=\frac{1}{d_F}\sum_{\rho}e^{ir_\rho}
\eeq
and do not depend explicitly on $\mu$. For SU($3$), we have
\beq
\label{eq:popolsu3}
 \ell(r,\mu)=\frac{1}{3}\left[e^{-i\frac{r_8}{\sqrt{3}}}+2e^{i\frac{r_8}{2\sqrt{3}}}\cos(r_3/2)\right].
\eeq

It is easy to check that the above expressions reproduce the SU($3$) results of Ref.~\cite{Reinosa:2014ooa}. Also, it is a simple exercise to verify that the leading-order expressions \eqn{eq_potq1loop}, \eqn{eq:V0f}, \eqn{eq:onelooppot}, \eqn{eq:Weiss0}, and \eqn{eq:popolsu3} satisfy all the symmetry properties derived in the previous section. To this aim, it is useful to note that the scalar products $u_i\cdot \rho_j$, $u_i\cdot \alpha_j$, and $e_i\cdot \alpha_j$  are all multiples of $2\pi$, whereas $e_1\cdot \alpha_j=2\pi/3\mod 2\pi$ and $e_2\cdot \alpha_j=-2\pi/3\mod 2\pi$.

\section{Vanishing chemical potential}
\label{sec:vanishing}

We now apply the above calculations to various situations of interest. We begin our discussion with the case $\mu=0$, where the lattice simulations are well under
control. We first discuss how the symmetry considerations of \Sec{sec:sym} constrain the background field potential at $\mu=0$. Then we study the phase structure 
of the theory as a function of the quark masses and we compare our findings with lattice results.

\subsection{Symmetries}
\label{sec:sym_mu=0}

A direct consequence of the relations \eqn{eq_muto-mu} and \eqn{eq_Zstar} at $\mu=0$ is that the generating
function is real if we choose the source and the background field components to be both real. Indeed, these relations imply 
\beq
\label{eq:kpon}
 Z(J,\Ab,0)=Z(-J,-\Ab,0)=Z^\star(J,\Ab,0).
\eeq
In particular, this guarantees that the average value $A$ of the gauge field at nonvanishing source, defined in \Eqn{eq:sourceAJA}, is real. This is important in the present setup where we eventually choose $\Ab$ such that  $A=\Ab$. Moreover, the above relations, together with Eqs.~\eqn{eq_polmuto-mu} and \eqn{eq_polconj} imply, for a real background field,
\beq
\label{eq:symdeplus}
 V(r,0)=V(-r,0)=V^\star(r,0)
\eeq
and
\beq
\label{eq:show}
 \ell(r,0)=\bar\ell(-r,0)=\ell^\star(-r,0).
\eeq
 In this case, the physical point is obtained by minimizing the background field potential.\footnote{The standard proof that the physical point corresponds to the absolute minimum of the potential is based on the convexity of $-\ln Z(J,\bar A,\mu)$ which is easily shown in the case where the action in the functional integral \eqn{eq_Z} is real. We note that this is not guaranteed in the usual FP quantization procedure since the determinant of the FP operator---obtained by integrating out the ghost and antighost fields---is not positive definite. We expect the situation in the present massive extension to be more favourable to a convex $-\ln Z(J,\bar A,\mu)$ since the mass term suppresses the large field configurations for which the FP operator develops negative eigenvalues. A more thorough investigation of these aspects goes beyond the scope of the present work and we postpone it to a future work (we stress that this is not particular to the present model but it is a general issue for background field methods). Here, we shall make the conservative assumption that in the cases where the background field potential is a real function of real variables, the physical point corresponds to an absolute minimum.} 

\begin{figure}[t]
\includegraphics[width=\linewidth]{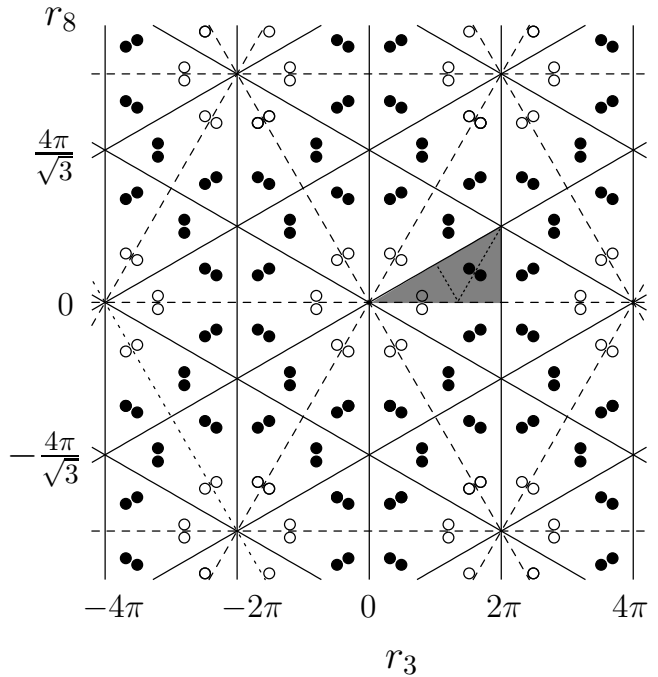}
\caption{ The diagram summarizes the symmetries of the
  background field potential at vanishing chemical potential in the plane
  $(r_3,r_8)$, with $r_{3,8}\in\mathds{R}$. The plain lines are mirror symmetry axes which reflect the global color symmetry and the background field 
  gauge invariance of the original action, see Eqs.~\eqn{eq:local1} and \eqn{eq:local2}. The latter induces the translation symmetries along the vectors $u_{1,2}$, defined in \Eqn{eq_vecs}. These are symmetries of the potential also at nonvanishing $\mu$. For $\mu=0$, charge conjugation invariance results in an inversion symmetry about the origin and the dashed lines thus also become mirror symmetry axes. The white dots are all equivalent by combination of these symmetries. The shaded triangle represents an elementary cell that is repeated all over the plane. In the pure gauge theory, the potential is also
  invariant under translations along the vectors $e_{1,2}$, defined in \Eqn{eq_vecs2}, such that the black and the white dots are all equivalent by
  symmetry. The elementary cell is reduced to a third of the shaded triangle, i.e., to any of the three subtriangles shown with dotted lines.}
\label{fig_sym_mu=0}
\end{figure}

The inversion symmetry \eqn{eq:symdeplus} implies that there are three more reflexion symmetry axes\footnote{These are symmetries of the potential, not exactly of the function $\ell(r,0)$,  which gets complex conjugated in the inversion through the origin; see \Eqn{eq:show}.} in the plane $(r_3,r_8)$ on top of those identified in \Sec{sec_color}. One is the axis $r_3=0$ and the two other are the lines which pass through the origin and which make an angle $\pm\pi/3$ with the $r_8=0$ axis. This is represented in \Fig{fig_sym_mu=0}. The origin and the points related to it by the translations described in Sec.~\ref{sec_perio}  now have the symmetry $C_{6v}$.

Using the symmetries of the potential, we can restrict the search for its minimum to the elementary cell depicted as a shaded triangle on \Fig{fig_sym_mu=0}. As discussed in Appendix~\ref{appeq:SSB}, the assumption that charge conjugation symmetry is not spontaneously broken  at $\mu=0$ implies that the absolute minimum of the background potential lies on the axis $r_8=0$ in this cell. This is indeed, what we find from the detailed investigation of the one-loop potential. 
Furthermore, Eqs.~\eqn{eq:show} and \eqn{eq_polreflexion} imply that $\ell(r,0)=\bar\ell(r,0)\in\mathds{R}$ on the axis $r_8=0$ so that the physical Polyakov loops, Eqs.~\eqn{eq_popoldef} and \eqn{eq_popoldefbar}, are equal and real,
\beq
 \ell(\mu=0)=\bar\ell(\mu=0)\in\mathds{R},
\eeq
as expected. Indeed, the fact that they are real is consistent with their standard interpretation in terms of the free energy of a static quark or antiquark \cite{Svetitsky:1985ye,Philipsen:2010gj}. The fact that they are equal simply tells that there is no distinction between the free energy of a quark and that of an antiquark at $\mu=0$, as a result of the charge conjugation symmetry.

\subsection{One-loop results}
\label{sec_res_zero_chem}

To analyze the phase diagram at $\mu=0$, we track the absolute minimum of the
one-loop background field potential computed in \Sec{sec:one_loop} as a function of temperature for different values of the quark masses. We consider the case of two degenerate flavors with mass $M_u$ and a third flavor with mass $M_s$. In the following, we shall either present our results in units of $m$ for dimensionful quantities, or compute dimensionless quantities that can be compared with lattice results. To get a rough idea of scales, a typical value used to fit lattice propagators at zero temperature for the SU($3$) Yang-Mills theory is $m\approx0.5$~GeV \cite{Reinosa:2014ooa,Tissier_11}.

\begin{figure}[t]
  \centering
  \includegraphics[width=.8\linewidth]{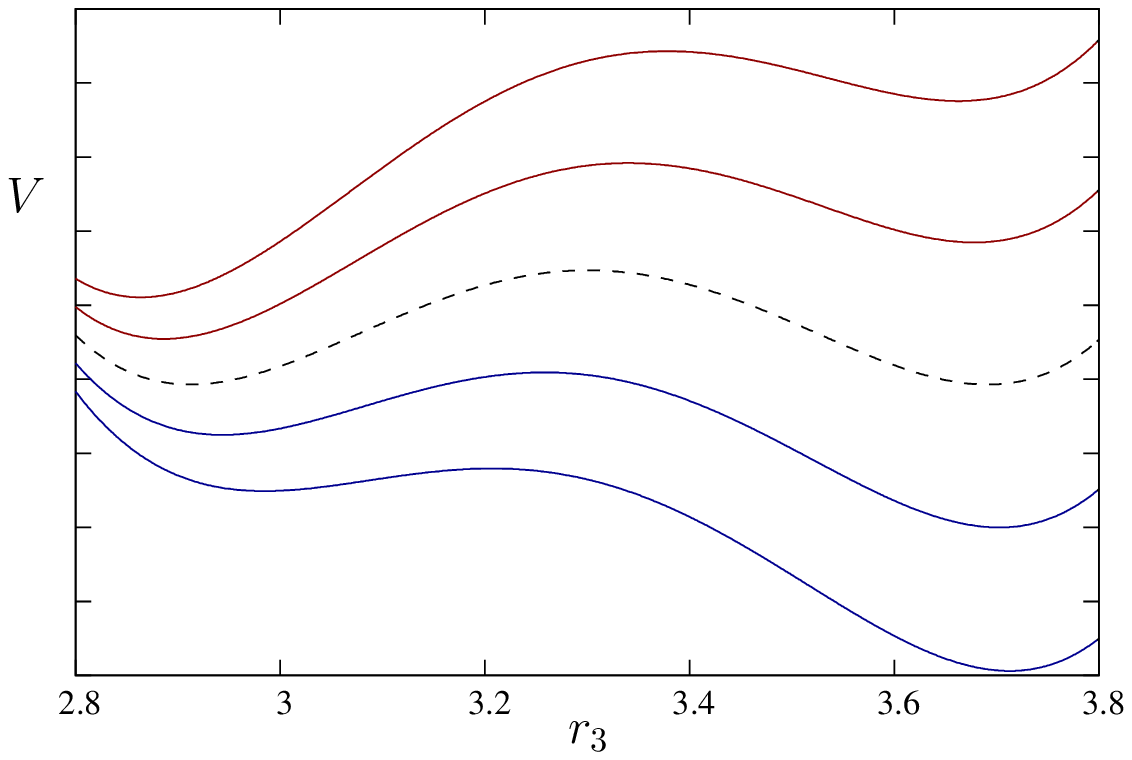}
  \includegraphics[width=.8\linewidth]{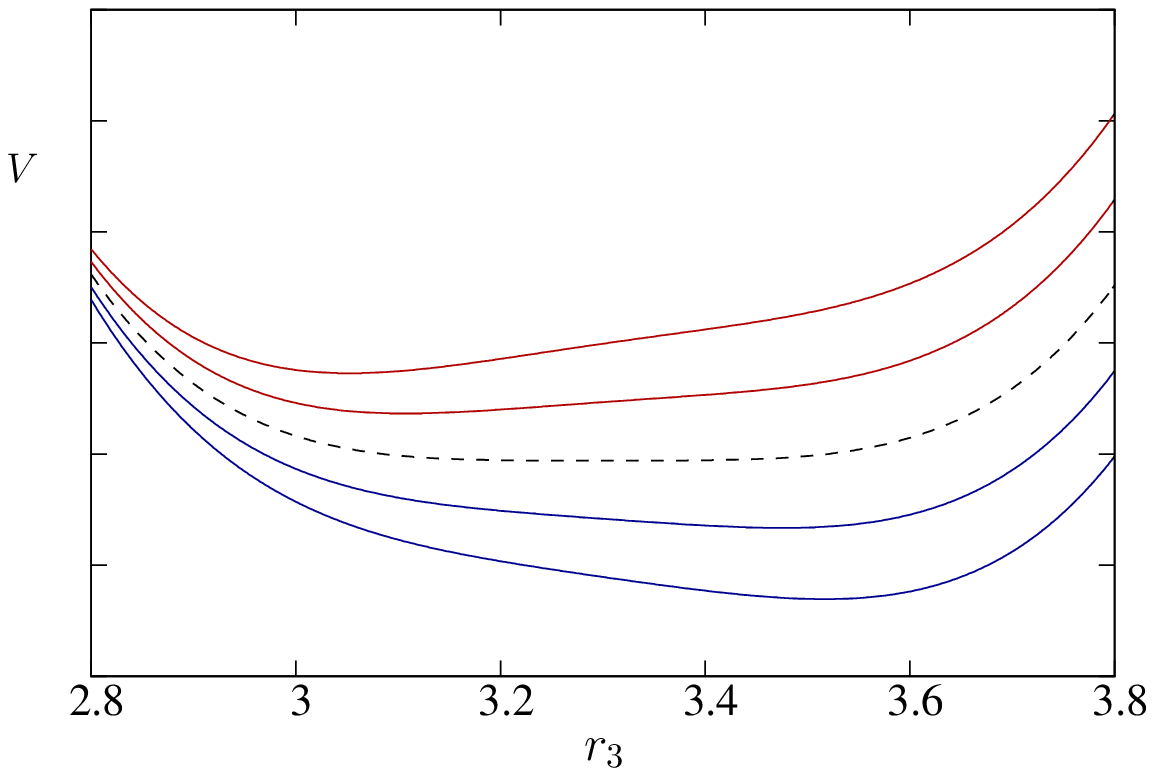}
  \includegraphics[width=.8\linewidth]{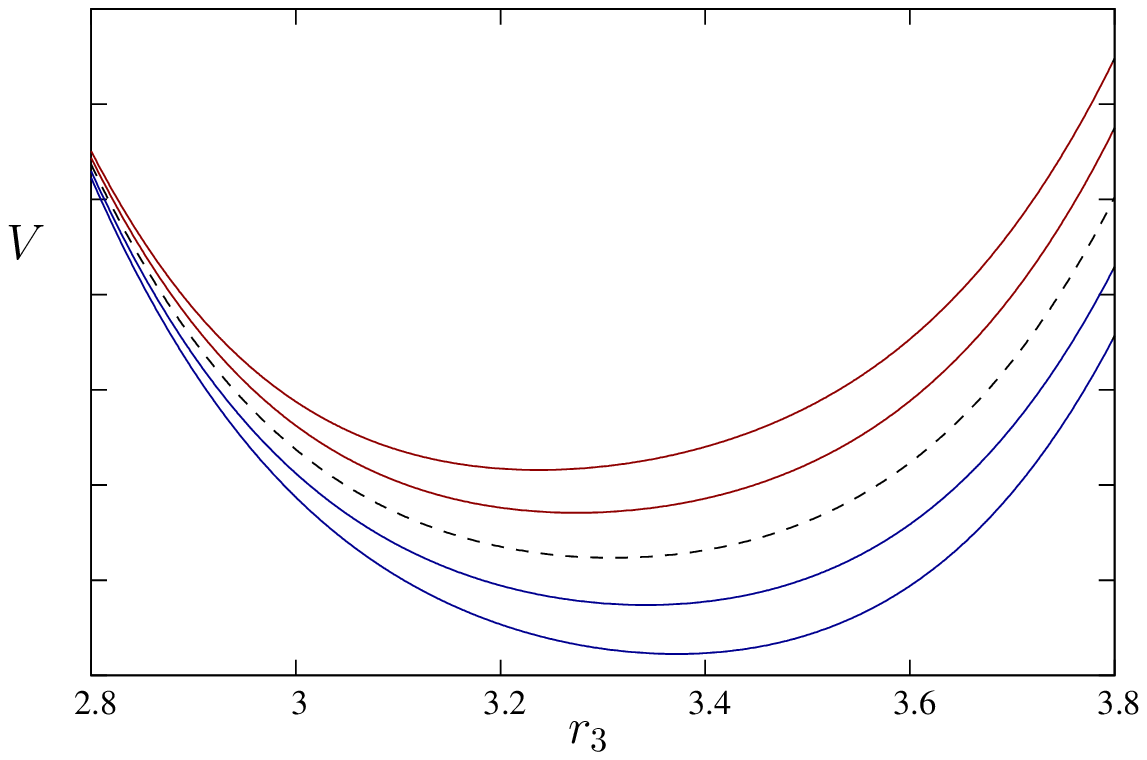}
  \caption{The background field potential (in arbitrary units) at $\mu=0$ as a function of $r_3$ for
    $r_8=0$ for different temperatures in the degenerate case $M_u=M_s=M$. Top
    figure: $M$ slightly larger than the critical value $M_c$; see
    Fig.~\ref{fig_columbia_mu_O}. Temperature increases for curves from bottom to top. The dashed line corresponds to the
    transition temperature. Middle
    figure: $M=M_c$ with the same conventions as
    for the top figure. Bottom figure: $M<M_c$. The dashed line is for the crossover
    temperature, where the curvature of the potential at the minimum is the smallest.}
  \label{fig_pot}
\end{figure}

\begin{figure}[t]
  \centering
  \includegraphics[width=.9\linewidth]{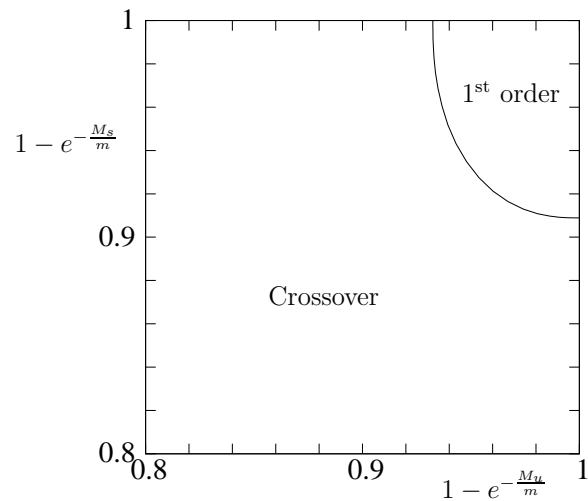}
  \caption{Columbia plot at $\mu=0$. In the upper
    right corner of the plane $(M_u,M_s)$, the phase transition is
    of first-order. In the lower left corner, the system presents a
    crossover. On the plain line, the system has a critical behavior. }
  \label{fig_columbia_mu_O}
\end{figure}

Depending on the values of the quark masses, we find different types
of behaviors when we change the temperature, as illustrated on \Fig{fig_pot}.  For large masses, the absolute minimum presents a finite jump at some transition temperature, signaling a first-order transition. Instead, for small masses, there is
always a unique minimum, whose location rapidly changes with temperature in some crossover regime. At the common
boundary of these two mass regions, the system presents a critical
behavior: there exists a unique minimum of the potential for all
temperatures, which however behaves as a power-law around some
critical temperature. The associated nonanalytic behavior of the
minimum of the potential as a function of temperature is a consequence
of the fact that, at the critical temperature, the first, second and
third derivatives in the $r_3$ direction vanish, a criterion which we used in order to determine the critical line in the Columbia plot of \Fig{fig_columbia_mu_O}. In the degenerate case $M_u=M_s$ we get, for the critical mass, $M_c/m=2.867$ and, for the critical temperature, $T_c/m=0.355$. We thus have $M_c/T_c=8.07$. This dimensionless ratio does not depend on the value of $m$ and can be directly compared to lattice results. For instance, the calculation of Ref.~\cite{Fromm:2011qi} yields, for $3$ degenerate quarks, $(M_c/T_c)^{\rm latt.}=8.32$. We obtain similar good agreement for different numbers of degenerate quark flavors, as summarized in Table~\ref{tab:mu0}. 

\begin{table}[h!]
  \centering
  \begin{tabular}{|c|c|c|c|c|c|}
\hline
$\,N_f\,$&$\,M_c/m\,$&$\,T_c/m\,$&$\,M_c/T_c\,$&$\left(M_c/T_c\right)^{\rm latt.}$&$\left(M_c/T_c\right)^{\rm matr.}$\\
\hline
1&2.395&0.355&6.74&7.22(5)&8.04\\
\hline
2&2.695&0.355&7.59&7.91(5)&8.85\\
\hline
3&2.867&0.355&8.07&8.32(5)&9.33\\
\hline
  \end{tabular}
  \caption{Values of the critical quark mass and temperature for $N_f=1,2,3$ degenerate quark flavors from the present one-loop calculation. The values of the dimensionless and parameter independent ratio $M_c/T_c$ are compared to the lattice results of Ref.~\cite{Fromm:2011qi} (before to last column). { For comparison we also show the values from the matrix model of Ref.~\cite{Kashiwa:2012wa} (last column).}}\label{tab:mu0}
\end{table}

We observe that the critical temperature is essentially unaffected by the presence of quarks. It is actually close to the one obtained in the present approach for the pure gauge SU($3$) theory \cite{Reinosa:2014ooa}. This is due to the fact that, for the typical values of $M/T$ near the critical line, the quark contribution to the potential is Boltzmann suppressed as compared to that of the gauge sector, as discussed in Appendix~\ref{appsec:approx}.

{ Finally, we compare our results for the ratio $M_c/T_c$ to those of Ref.~\cite{Kashiwa:2012wa} based on matrix models. We also mention that recent calculations in the Dyson-Schwinger approach \cite{Fischer:2014vxa} yield values of the ratio $M_c/T_c$ that are systematically smaller than the ones obtained on the lattice. The origin of this discrepancy remains to be clarified.}
 
\section{Imaginary chemical potential}
\label{sec:im}

The study of the QCD phase diagram for imaginary chemical potential is of great interest. There is no sign problem in this case and lattice simulations can be performed in a standard way \cite{de Forcrand:2003hx,D'Elia:2002gd}. Thermodynamic potentials at real chemical potential can then be obtained, in principle, by analytic continuation. In practice, however, this is restricted to small values $\mu/T\lesssim1$ because the theory at imaginary chemical potential has a nontrivial phase structure with nonanalytical behaviors. The phase diagram at imaginary chemical potential is also interesting in itself as it presents {\it bona fide}
properties of QCD \cite{Roberge:1986mm,deForcrand:2010he}. In this section, we describe
the predictions of our one-loop calculation for a purely imaginary chemical potential $\mu=i\mu_i$. As in the previous case, we begin the discussion by analyzing the consequences of the symmetry properties discussed in \Sec{sec:sym}.
 
\subsection{Symmetries}

As before, we seek conditions on the source $J$ and the background field $\Ab$ such that the identification $A=\Ab$, with $A$ defined in \Eqn{eq:sourceAJA} is consistent. The relations \eqn{eq_muto-mu} and \eqn{eq_Zstar} imply that the generating function is real if we choose both the source and background field to be real. We have, in that case,
\beq
  Z(J,\Ab,i \mu_i)=Z(-J,-\Ab,-i\mu_i)=Z^\star(J,\Ab,i \mu_i).
\eeq
The average field $A$ of \Eqn{eq:sourceAJA} is thus real and can be consistently identified with $\Ab$. The background field potential $V(r,i\mu_i)$ is also real, the Polyakov loop functions are related by complex conjugation, $\bar\ell(r,i\mu_i)=\ell^\star(r,i\mu_i)$, and so are the physical Polyakov loops:
\beq
 \bar\ell(i\mu_i)=\ell^\star(i\mu_i).
\eeq
The Polyakov loop is complex for generic values of $\mu_i$ and its argument actually plays the role of an order parameter for the various transitions described below \cite{Roberge:1986mm}. Finally, the background field potential being a real function of real variables  as in the case $\mu=0$, we adopt the same prescription as before to choose the background field for the evaluation of physical observables, namely we minimize the potential $V$.

\begin{figure}[t]
  \centering
  \includegraphics[width=\linewidth]{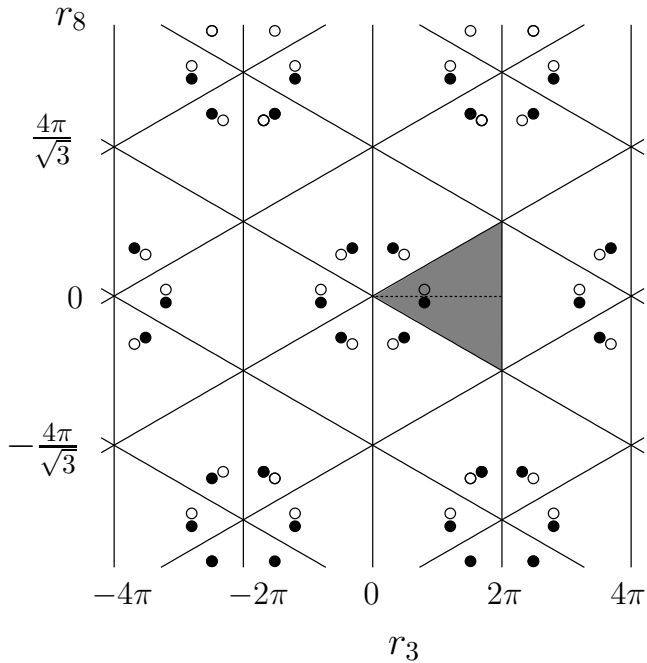}
  \caption{The white (black) dots are equivalent up to symmetry
    transformations that exist for a generic chemical potential. In
    particular, the system is invariant under reflexion with respect
    to the plain lines. The shaded equilateral triangle corresponds to the
    elementary cell that is repeated all over the plane. For imaginary chemical potential $\mu=i\mu_i$, 
    there exists specific values, $\mu_i=n\pi /\beta$, with $n\in\mathds{Z}$, for which the
    theory is invariant under inversion about the origin and the white and black dots are equivalent. The shaded triangle is
    symmetric under reflexion about the dashed vertical line so that
    the elementary cell is now half of the shaded equilateral
    triangle. Finally, the symmetries of the effective potential imply that the elementary triangle of the theory at imaginary chemical potential $\mu_i\pm2\pi/(3\beta)$ can be obtained from that of the theory at $\mu_i$ by a rotation of $\pm2\pi/3$ around its center.}
  \label{fig_sym_imag}
\end{figure}

The symmetry properties of the potential $V(r,i\mu_i)$ for a generic value of $\mu_i$
are summarized in \Fig{fig_sym_imag}. The white dots are all images of each others by the global color symmetries \eqn{eq_polreflexion} and the gauge transformations \eqn{eq_translat} and are thus all physically equivalent. The same holds for the black dots but, contrarily to the case $\mu_i=0$ discussed in the previous section, the black and white dots are not equivalent. The elementary cell, to which one can restrict the analysis, is thus twice as large as that of the previous section. It can be chosen as the shaded equilateral triangle shown in \Fig{fig_sym_imag}. Observe that the origin [and the equivalent points obtained by the translations \eqn{eq_translat}] have the symmetry $C_{3v}$. 

Remarkably, the elementary cell has an extra symmetry when the imaginary chemical potential $\mu_i$ is a multiple of $\pi/(3\beta)$. Using the general periodicity property $V(r,\mu)=V(r,\mu+2i\pi/\beta)$, one can reduce the discussion to the interval $0\le\mu_i<2\pi/\beta$. The case $\mu_i=0$ has been discussed before and possesses a $C_{6v}$ symmetry around the origin [and all the equivalent points obtained by the translations \eqn{eq_translat}]. This is also true for $\mu_i=\pi/\beta$. Indeed, using Eq.~(\ref{eq_muto-mu}) and the $2\pi/\beta$-periodicity in $\mu_i$, we have
\beq
 V(r,i\pi/\beta)=V(-r,-i\pi/\beta)=V(-r,i\pi/\beta).
\eeq
This corresponds to an inversion about the origin and we thus retrieve the same set of symmetries as in the case $\mu_i=0$, depicted in \Fig{fig_sym_mu=0}.  Finally, the cases $\mu_i=2n\pi/(3\beta)$ and $\mu_i=(2n+1)\pi/(3\beta)$, with $n$ integer, can be obtained from the cases $\mu_i=0$ and $\mu_i=\pi/\beta$, respectively, by using the relation \eqn{eq_center}. The symmetries are similar to those depicted in \Fig{fig_sym_mu=0} where, however, the point with $C_{6v}$ symmetry is not the origin anymore but one of the two other vertices of the elementary triangle [and their images by the translations \eqn{eq_translat}]. For $\mu_i=\pi/(3\beta)$ and $\mu_i=4\pi/(3\beta)$, this is the upper vertex, located at $(2\pi,2\pi/\sqrt 3)$, whereas for $\mu_i=2\pi/(3\beta)$ and $\mu_i=5\pi/(3\beta)$, it is the lower vertex, located at $(2\pi,-2\pi/\sqrt 3)$. Using the axes of symmetry corresponding to global color transformations \eqn{eq_Vreflexion} (the plain lines in \Fig{fig_sym_mu=0}), one sees that the elementary triangle is rotated by $\pm 2\pi/3$ each time the imaginary chemical potential is shifted by $\pm 2\pi/(3\beta)$. The corresponding Polyakov loop acquires a phase $\pm2\pi/3$.

\subsection{One-loop results}
\label{sec_res_mu_im}

\begin{figure}[t]
  \centering
  \includegraphics[width=.8\linewidth]{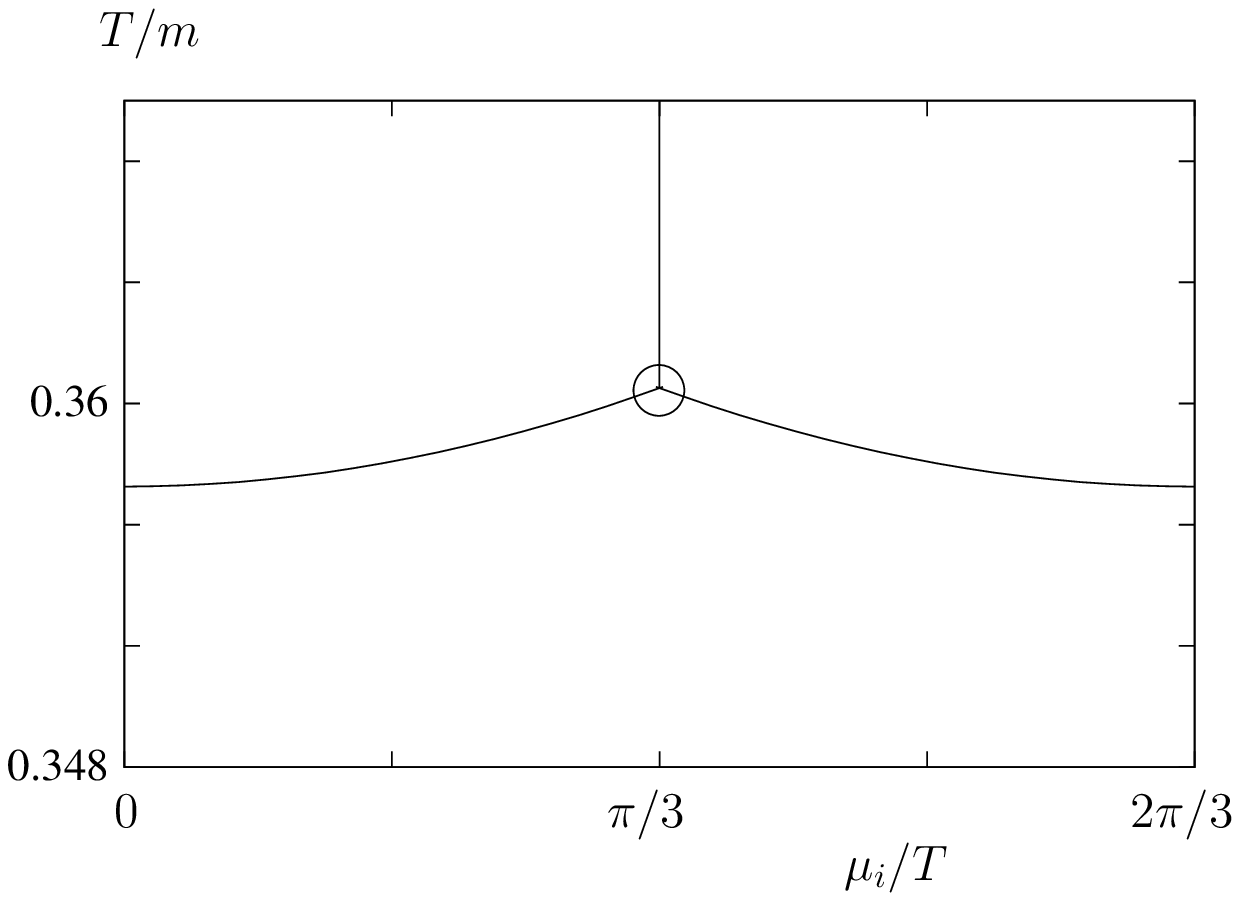}
  \includegraphics[width=.8\linewidth]{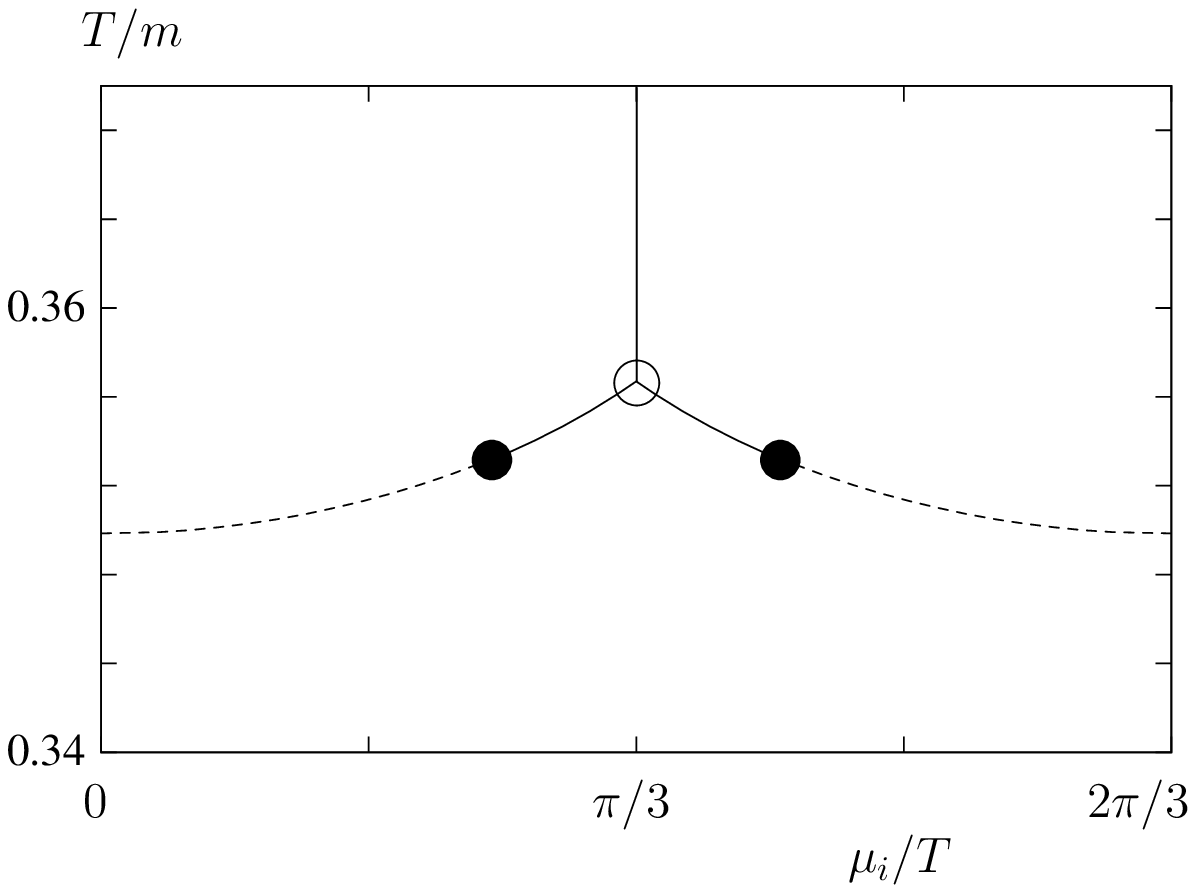}
  \includegraphics[width=.8\linewidth]{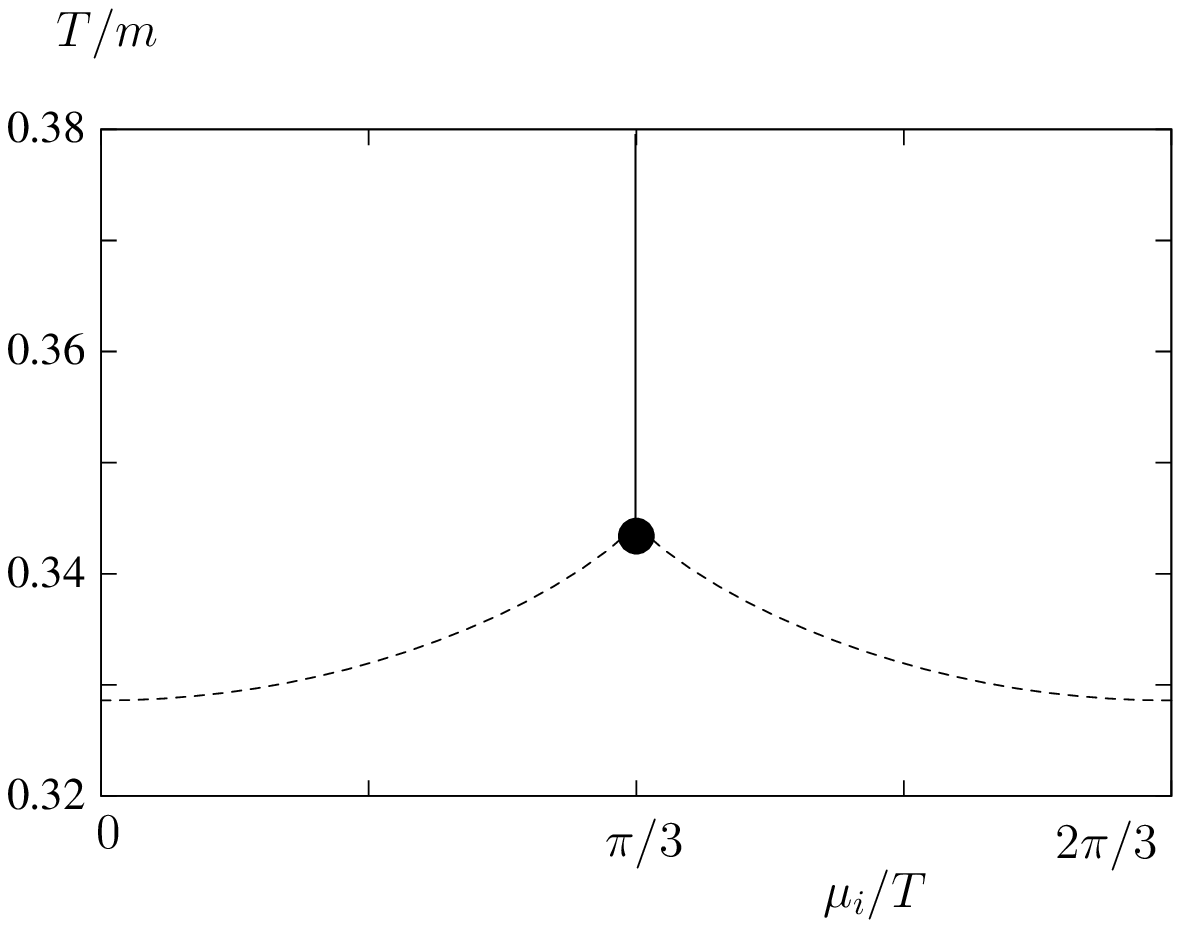}
  \caption{ Phase diagram in the plane $(\mu_i/T,T)$ for different values of the degenerate quark mass $M$. Top panel: $M=3m$ is larger that the critical mass at vanishing chemical potential $M_c(0)\simeq2.8m$. Middle panel: $M=2.6m$ is below $M_c(0)$ but larger than the second critical mass $M_c(i\pi T/3)\simeq2.2m$; see text. Bottom panel: $M=2m$ is smaller that $M_c(i\pi T/3)$. Plain lines correspond to first-order phase transitions,
    dashed lines to crossovers, black dots represent $Z_2$ critical
    points, and the empty circles are triple points.}
  \label{fig_phase_diag}
\end{figure}

We now present our results for the phase diagram in the plane
$(\mu_i/T,T)$. Using the property \eqn{eq:partitionfunc}, it is sufficient to study the interval $0\le\mu_i/T\le2\pi/3$ and the phase diagram is symmetric around $\mu_i/T=\pi/3$. For simplicity, we consider the case of
three degenerate quarks with $M_u=M_s=M$. For any mass larger
than $M_c(\mu=0)\simeq 2.8m$, such that the transition at
vanishing chemical potential is first-order (see Fig.~\ref{fig_columbia_mu_O}), we
find that the transition persists at nonvanishing $\mu_i$, as depicted in the top panel of \Fig{fig_phase_diag}. Two first-order phase transition lines join at the symmetry axis $\mu_i/T=\pi/3$. Moreover, for sufficiently large temperatures, there exists a first-order phase transition
across $\mu_i/T=\pi/3$, where the argument of the Polyakov loop has a finite jump, signaling the spontaneous breaking of the Roberge-Weiss symmetry\footnote{When the Roberge-Weiss symmetry is not broken, \Eqn{eq:ellsym} implies that $e^{-i\pi/3}\ell(i\pi T/3)\in\mathds{R}$ so that ${\rm Arg}\,\ell(i\pi T/3)=\pi/3$.} \cite{Roberge:1986mm}. { In the same range of temperatures, we also observe that the reflection symmetry of the potential at $\mu_i/T=\pi/3$ with respect to the median passing through the vertex $(2\pi,2\pi/\sqrt{3})$ is spontaneously broken. As $\mu_i/T$ changes from $\pi/3^-$ to $\pi/3^+$, the minimum has a finite jump from one to the other side of this median. This leads to a cusp in the partition function at $\mu_i/T=\pi/3$, as anticipated in \cite{Roberge:1986mm}.} The three first-order lines meet at a triple point, where the system can coexist in three different phases characterized by different values of the argument of the Polyakov loop or, equivalently at one-loop order, by the value of the background field in the $r_8$-direction at the minimum of the potential. 

\begin{figure}[t]
  \centering
  \includegraphics[width=.9\linewidth]{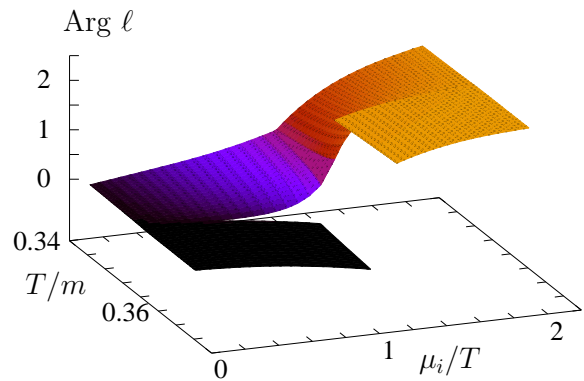}
  \caption{Argument of the Polyakov loop in the $(\mu_i/T,T)$ plane for a degenerate quark mass $M$ in the range $M_c(i\pi T/3)<M<M_c(0)$, corresponding to the middle panel of \Fig{fig_phase_diag}. first-order transitions appear as a discontinuities in this quantity. The discontinuity vanish at the $Z_2$ critical points. At the triple point on the line $\mu_i/T=\pi/3$ , the argument of the Polyakov loop can take three different values. }
  \label{fig_arg_pol}
\end{figure}

For $M=M_c(\mu=0)$, the transition at vanishing chemical potential is second order and there appears, in the $(\mu_i/T,T)$ phase diagram, a couple of $Z_2$ critical points which terminate the lines of first-order transitions described above at $\mu_i/T=0$ and $\mu_i/T=2\pi/3$. Decreasing the mass $M$ further, these critical points penetrate deeper in the phase diagram towards $\mu_i/T=\pi/3$, as shown in the middle panel of  Fig.~\ref{fig_phase_diag}. At a critical value $M_c(\mu=i\pi T/3)\simeq 2.2m$, the  two $Z_2$ critical points merge at the symmetric point $\mu_i/T=\pi/3$ and give rise to a tricritical point which terminates the vertical line of first-order transition \cite{deForcrand:2010he}. The horizontal lines of first-order transitions for $\mu_i/T\neq\pi/3$ have completely disappeared and are replaced by crossovers. For $M<M_c(i\pi T/3)$, the picture is the same with, however, the tricritical point ending the first-order transition line at $\mu_i/T=\pi/3$ replaced by a $Z_2$ critical point, as shown in the lower panel of \Fig{fig_phase_diag}.  As a further illustration, we display in Fig.~\ref{fig_arg_pol} the argument of the Polyakov loop in the $(\mu_i/T,T)$ plane for the intermediate mass case, $M_c(i\pi/3)<M<M_c(0)$. Similar plots can be made in the other cases as well. { We mention that similar results were obtained using effective matrix models in Ref.~\cite{Kashiwa:2013rm}.}

The tricritical point at $\mu_i/T=\pi/3$ and $M=M_c(i\pi T/3)$ features particular scaling laws which constrain the $\mu_i$-dependence of the critical quark mass and temperature, $M_c$ and $T_c$, at the $Z_2$ critical points. If the scaling window is broad enough, this may put constraints on the phase diagram away from the tricritical point and possibly for real chemical potential as well \cite{deForcrand:2010he}. Our results  can be qualitatively described by the following model potential
\begin{equation}
  \label{eq_pot_toy}
  U(\phi)=\frac 16\phi^6+\frac a4\phi^4+\frac b2\phi^2-h\phi
\end{equation}
where $a\propto M_c(i\pi T/3)-M$, $b$ is a decreasing function of the temperature, and the $Z_2$-breaking term $h\propto
(\pi/3)^2-(\mu_i/T)^2$ characterizes the explicit breaking of the Roberge-Weiss symmetry, that is, the distance to the symmetry axis $\mu_i/T=\pi/3$. Here,  $\phi$ parametrizes the curve in the $(r_3,r_8)$ plane along which the minimum of the potential $V(r,i\mu_i)$ moves. Equation \eqn{eq_pot_toy} is the minimal model for describing the
tricritical scaling. For $h=0$ and $a<0$, it describes a first-order transition with a jump in $\phi$ at the minimum $\propto\sqrt{-a}$. Instead, the phase transition is continuous with a $Z_2$ critical point for for $a>0$ and a tricritical point in the
limiting case $a=0$. This qualitatively describes the various cases depicted in \Fig{fig_phase_diag} on the axis $\mu_i/T=\pi/3$.

\begin{figure}[t]
  \centering
  \includegraphics[width=.9\linewidth]{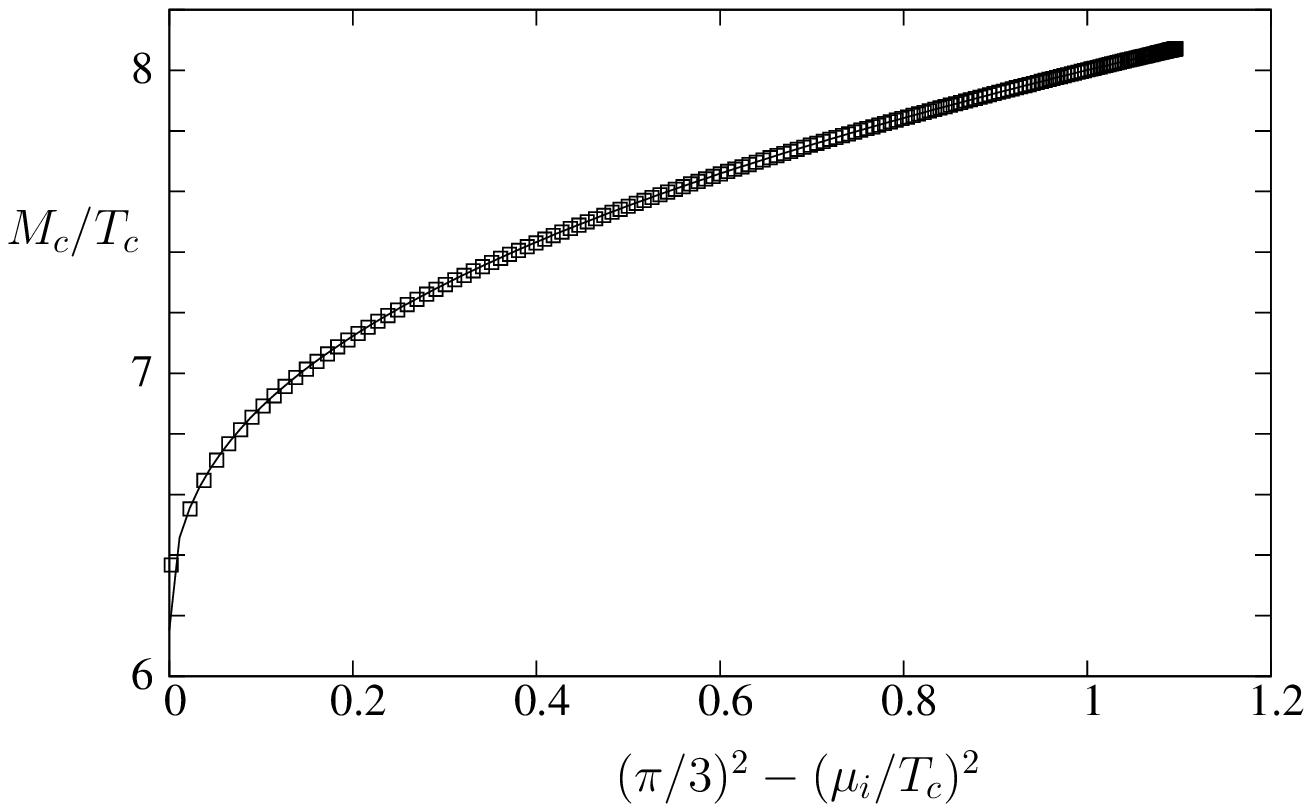}\vspace{.5cm}
  \includegraphics[width=.9\linewidth]{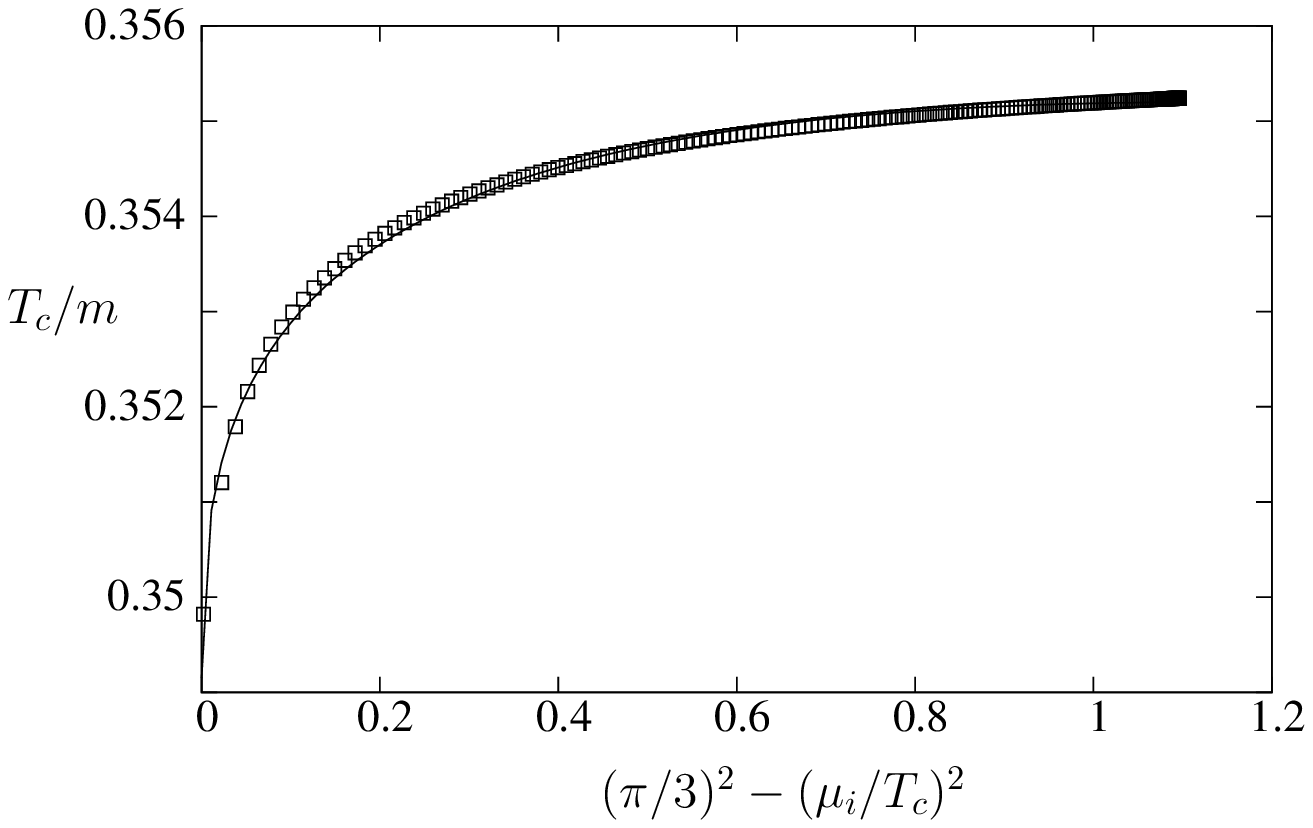}
  \caption{Top panel: The quark mass at the critical point as a
    function of $x=(\pi/3)^2-(\mu_i/T)^2$. The white squares are the results of the present one-loop calculation and the plain line is the fit $6.15+1.85\,x^{2/5}$.  Bottom panel: The temperature in units of $m$ at
    the critical point as a function of $x$. The plain
    line is the fit $0.35+0.012\,x^{2/5}-0.0055\,x^{4/5}$.}
  \label{fig_mass_tricrit}
\end{figure}

Now, for $h\neq0$, the potential
(\ref{eq_pot_toy}) predicts a crossover as a function of temperature for $a\ge0$ and either a first-order transition for sufficiently large negative $a$ or a crossover for small negative $a$. This corresponds to the various cases of \Fig{fig_phase_diag} away from the axis $\mu_i/T=\pi/3$. The critical point in the middle panel is the limiting case between the first-order transition and the crossover for $a<0$. The corresponding critical values scale as $a_c\sim h^{2/5}$ and $b_c\sim h^{4/5}$. This directly translates to the scaling law for the critical quark mass and the critical temperature.  Such scalings are well reproduced by our results, as shown in
Fig.~\ref{fig_mass_tricrit}. The scaling of the critical quark mass directly follows that of the parameter $a$ in the above model. However, the parameter $b$ is not only a function of the temperature but also depends on the mass. As a consequence, the scaling of the critical temperature has both a contribution in $h^{2/5}$
and a contribution in $h^{4/5}$. As mentioned before, such scaling relations can be extended to arbitrary chemical potential by replacing $\mu_i\to -i\mu$. In particular, they can be compared to direct calculations of the phase diagram at real chemical potential, as discussed in the next section.

It is also interesting to compare our results to lattice calculations. Following Ref.~\cite{deForcrand:2010he}, we write
\beq
\label{eq:tricsca}
 \frac{M_c(\mu)}{T_c(\mu)}= \frac{M_{\rm tric.}}{T_{\rm tric.}}+K\left[\left(\frac{\pi}{3}\right)^2+\left(\frac{\mu}{T}\right)^2\right]^{2/5}.
\eeq
A fit of our results at $\mu=i\mu_i$ yields $M_{\rm tric.}/T_{\rm tric.}=6.15$ and $K=1.85$, to be compared with the lattice result of Ref.~\cite{Fromm:2011qi}, $(M_{\rm tric.}/T_{\rm tric.})^{\rm latt.}=6.66$ and $K^{\rm latt.}=1.55$ for $3$ degenerate quark flavors.\footnote{We mention that the value $K^{\rm latt.}=1.55$ of Ref.~\cite{Fromm:2011qi} corresponds to the case $N_f=1$. Although these authors do not mention the corresponding value for $N_f=3$, we observe that keeping the same value gives a reasonable estimate of the ratio $M_c/T_c$ at $\mu=0$ quoted in that reference.} As for the case $\mu=0$, we note that the values of the fitting parameters obtained in the recent Dyson-Schwinger calculation of Ref.~\cite{Fischer:2014vxa} are systematically smaller than the one obtained here and on the lattice. Again, the origin of this discrepancy remains to be elucidated.

\section{Real chemical potential}
\label{sec:re}

The case of real chemical potential is where the lattice methods suffer from a severe sign problem because the QCD action in the functional integral measure is not real. As we shall discuss below, this also affects continuum approaches, although in a much less severe way. As before, we first discuss the symmetries of the problem and the issue of choosing the source and background field in a consistent way. This is where the remnant of the sign problem appears. We then present the results of our one-loop calculation.

\subsection{Symmetries}
\label{sec_sym_mureal}

The main difficulty that we have to face at real chemical potential is the fact that the generating functional \eqn{eq_Z} is not real in general; see \Eqn{eq_Zstar}. As
a consequence, the average value $A$ of the gluon field, obtained from \Eqn{eq:sourceAJA}, is typically not real and one needs to consider complex background fields\footnote{With a complex background $\bar A$ the meaning of the shift (\ref{eq:Aba}) under the path integral needs to be reconsidered. In fact there is no contradiction because, strictly speaking, in the derivation of the loop expansion formula, the shift needs to be done after the original (real) contours of integration for the components of the gauge field have been deformed to pass through the relevant saddle point. At this level, the variables of integration are {\it a priori} complex and (\ref{eq:Aba}) is a mere complex change of variables for contour integrals.} (and possibly complex sources) in order to be able to identify $A=\Ab$ in a consistent way. However,  one has then to deal with a potential $V(r,\mu)$ which is a complex function of complex variables. Not only is the problem more intricate but it is not even clear what is the correct procedure to identify the physical value of the background field needed to compute observables (which should end up being real despite the fact that the extremum of $V(r,\mu)$ is possibly complex). For instance, these do not correspond in an obvious way to the absolute minimum of the potential anymore and might instead be saddle points. In that case, it is not clear which saddle point is the preferred one.

The problem is greatly simplified---although not completely solved---by choosing the source and background field as
\beq
 J=\left(\begin{tabular}{c}$J^3$\\$iJ^8$\end{tabular}\right)\,,\quad \Ab=\left(\begin{tabular}{c}$\Ab^3$\\$i\Ab^8$\end{tabular}\right).
\eeq
with $J^{3,8}$ and $\Ab^{3,8}$ real, such that\footnote{A similar proposal has been made in Ref.~\cite{Nishimura:2014kla}, where the authors also propose to search for saddle points of the effective potential $V(r_3,ir_8,\mu)$ in the plane $(r_3,r_8)$; see below.}
\beq
 R_1J=-J^\star\quad{\rm and} \quad R_1\Ab=-\Ab^\star,
\eeq
where $R_1$ is the reflexion with respect to the $8$-axis, defined in \Eqn{eq_R1}. Using Eqs.~\eqn{eq_Zstar} and \eqn{eq_reflexion}, one sees that the generating function is real,\footnote{By using the color
  transformations of Sec.~\ref{sec_color}, one can find other ways of
  having a real generating function where both  $r_3$ and $r_8$ are
  complex. These are however equivalent to the choice considered in
  the main text and lead to identical physical predictions. We have also investigated the case where both $r_3$ and $r_8$ are purely imaginary, for which $V(r,\mu)$ is also real. However, the one-loop expression of the potential does not make sense in that case.}
\beq
  Z[J,\Ab,\mu]=Z(R_1J,R_1\Ab,\mu)=Z^\star(J,\Ab,\mu).
\eeq
This has two important consequences. First, the averaged gluon field \eqn{eq:sourceAJA} is of the form $A=(A^3,iA^8)$ with $A^{3,8}$ real, which guarantees that the identification $A=\Ab$ is consistent. Second, it implies that the effective potential is a real function of the real variables $r_{3,8}=\beta g\Ab^{3,8}$:
\beq
 V(r,\mu)\equiv V(r_3,ir_8,\mu)=V^\star(r_3,ir_8,\mu).
\eeq
Note that the potential is not invariant under translations and reflexions in the (imaginary) $r_8$ direction but these symmetries are still valid in the $r_3$ direction. Consequently, it is sufficient to study the band $r_3\in[0,2\pi]$.

\begin{figure}[t]
  \centering
  \includegraphics[width=.8 \linewidth]{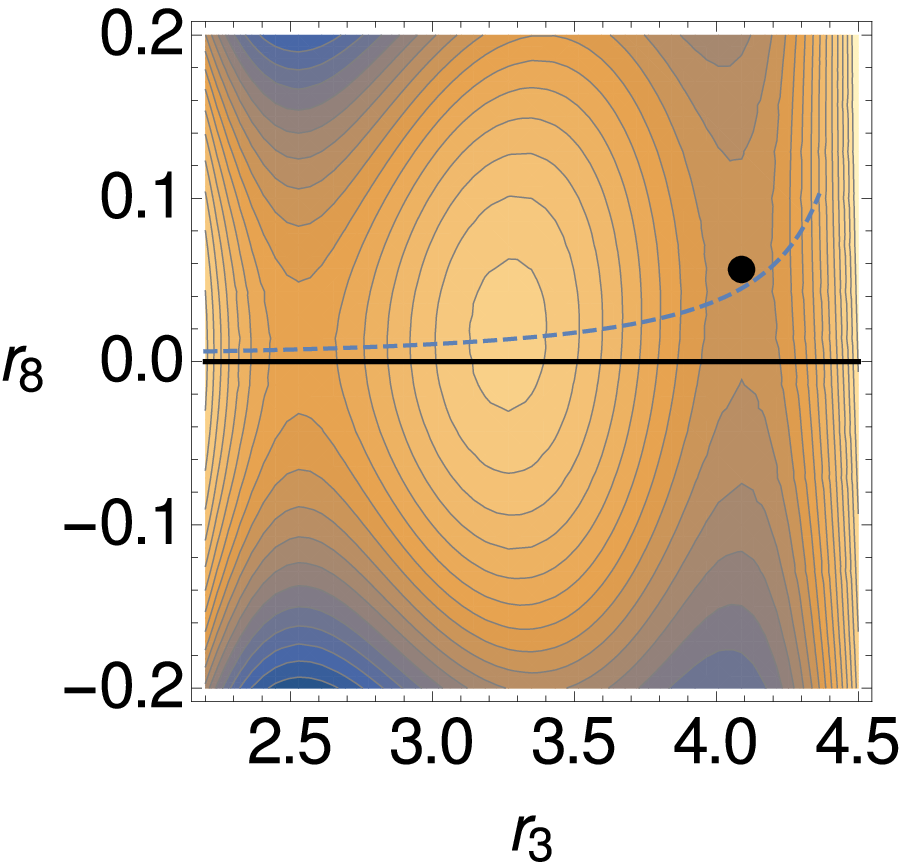}\vspace{.5cm}
  \includegraphics[width=.8 \linewidth]{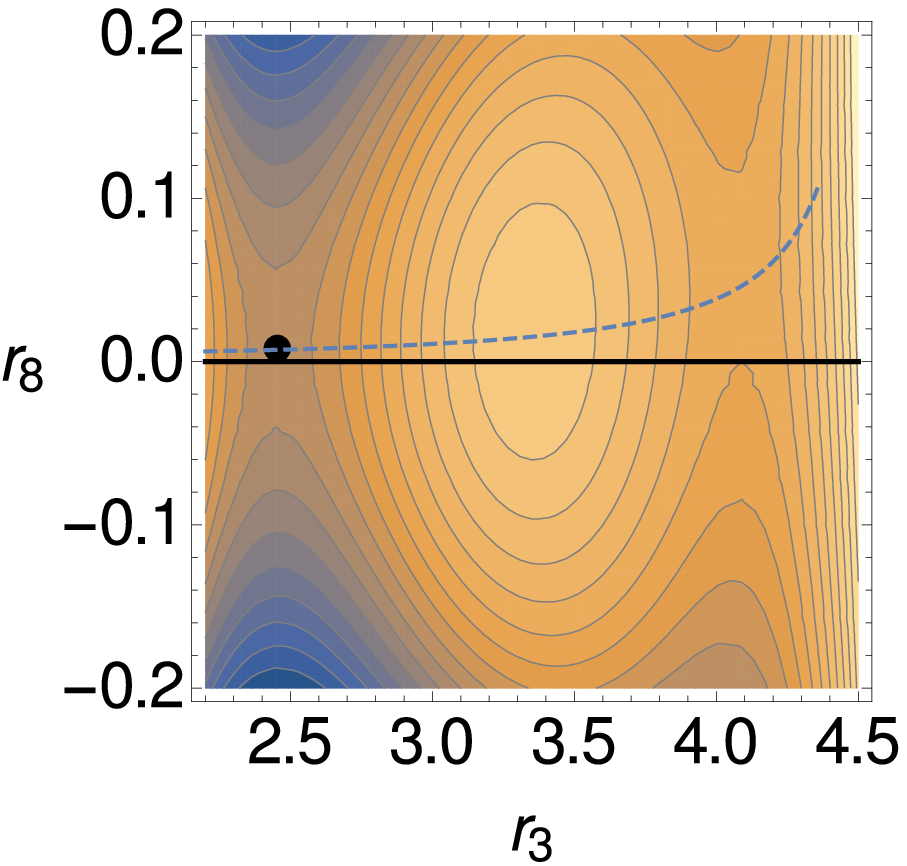}
  \caption{Contour plots of the background field potential $V(r_3,ir_8,\mu)$ in the plane $(r_3,r_8)$ for $M_u=M_s=4m$ and $\mu=0.6m$ for $T<T_c$ (top) and $T>T_c$ (bottom), with $T_c=0.36m$. The saddle point where the potential is the lowest is indicated by the black dot. We clearly see the jump at $T_c$ and the fact that the saddle points sit at  $r_8\neq0$. The dashed curve is an approximate analytic expression of the position of the saddle point in the $r_8$-direction as a function of its $r_3$ coordinate; see Appendix~\ref{appsec:approx}.}
  \label{fig_saddles}
\end{figure}

Now comes the question of selecting the correct extremum. We first observe that in the case
$\mu=0$, both $V(r_3,r_8,0)$ and $V(r_3,ir_8,0)$ are real for real $r_{3,8}$. In the former case, the relevant extrema are the absolute minima of the potential, which lie on the $r_8=0$ axis, as discussed in \Sec{sec:sym_mu=0}. It is clear that if we now change $r_8\to ir_8$, the curvature in the $r_8$ direction changes sign and these minima turn into saddle points. For $\mu\neq0$ not too large, we can follow these saddle points, which continuously move away from the $r_8=0$ axis. Which one to choose as the physical point is, however, not clear. Unlike in the case of a real action, we have no first principle argument for preferring a particular extremum. This can be seen as a remnant of the sign problem of lattice simulations; see \cite{Marko:2014hea} for a similar discussion in the context of a charged scalar field. Even if the complex nature of the action is not an obstacle to perform calculations using continuum approaches, it leads to ambiguities in selecting the physical solution out of the various extrema. Here, we always choose (arbitrarily) the saddle point for which the real function $V(r_3,ir_8,\mu)$ is the lowest. Although this seems reasonable at small $\mu$ by continuity with the case $\mu=0$, we have no guarantee that this is a valid procedure---if at all---for all values of $\mu$.  

Let us finally mention that with the choice of background field coordinates considered here, the Polyakov loops are always real, as follows from the relations \eqn{eq_polconj} and \eqn{eq_polreflexion}:
\beq
  \ell(r,\mu)=\ell(R_1r,\mu)=\ell^\star(r,\mu)
\eeq
and similarly for $\bar\ell(r,\mu)$, where $r=(r_3,ir_8)$. As for the case $\mu=0$, the fact that the Polyakov loops are real is consistent with their interpretation in terms of the free energy of static quarks or antiquarks. In general, $\ell(r,\mu)\neq\bar\ell(r,\mu)$ at $\mu\neq0$, which simply reflects the fact that the free energy of a static quark differ from that of its antiparticle at finite $\mu$. We note though that at tree-level we have the additional relation $\ell(r,\mu)=\bar\ell(-r,\mu)$ [see \Eqn{eq:popolsu3general}] so that the two functions coincide on the axis $r_8=0$. There is, however, neither any reason of symmetry nor physical argument for this to be the case and we do not expect it to be true at higher orders.

\subsection{One-loop results}

\begin{figure}[t]
  \centering
  \includegraphics[width=.9 \linewidth]{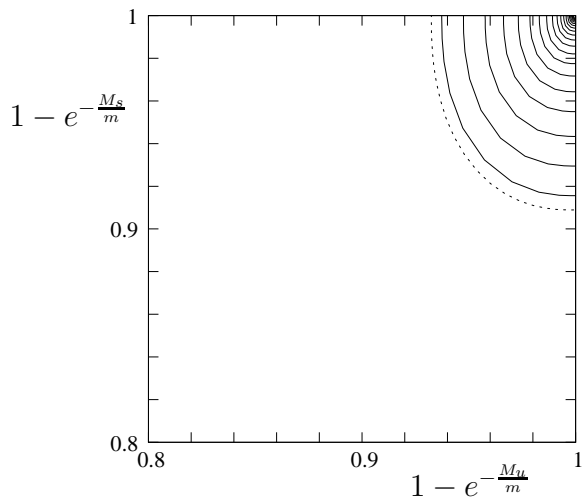}
  \caption{Columbia plot for real chemical potential. The first-order
    region contracts with increasing $\mu$. The dotted
    line corresponds to $\mu=0$. Successive plain lines correspond to a chemical
    potential increased by steps $\delta\mu=0.1$~GeV.}
  \label{fig_columbia_mu_real}
\end{figure}

We concentrate here on the determination of the critical line in the $(M_u,M_s)$ plane. We perform a similar analysis as in \Sec{sec:vanishing}, but this time identifying saddle points. A typical situation is illustrated in \Fig{fig_saddles}, where we see the jump of the deepest saddle point at the first-order transition. We recover the $\mu=0$ results and we find that the region of first-order transition shrinks towards large quark masses when the chemical potential increases, as shown in \Fig{fig_columbia_mu_real}.

\begin{figure}[t]
  \centering
  \includegraphics[width=.9 \linewidth]{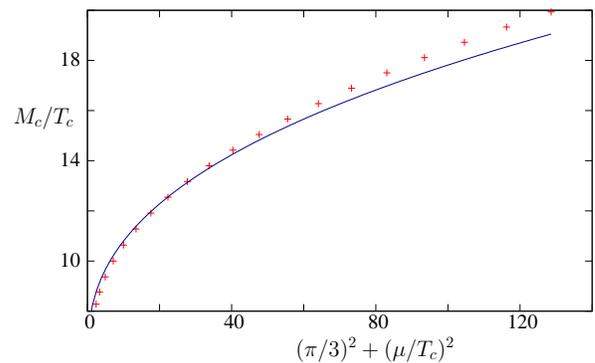}
  \caption{The critical mass $M_c(\mu)/T_c(\mu)$ as a function of $x=(\pi/3)^2+(\mu/T)^2$ for the case of degenerate quark masses $M_u=M_s$. The crosses are the result of the direct calculation. The line is obtained from the tricritical scaling law \eqn{eq:tricsca}; see also Fig.~\ref{fig_mass_tricrit}.}
  \label{fig:tricscalingmureal}
\end{figure}

We plot the $\mu$ dependence of the critical quark mass in the degenerate case $M_u=M_s$ in \Fig{fig:tricscalingmureal} and we compare it with the expectation from the tricritical scaling \eqn{eq:tricsca} extrapolated from the imaginary chemical potential region. We see that this describes well the data at real chemical potential.

\begin{figure}[t]
  \centering
\includegraphics[width=.9\linewidth]{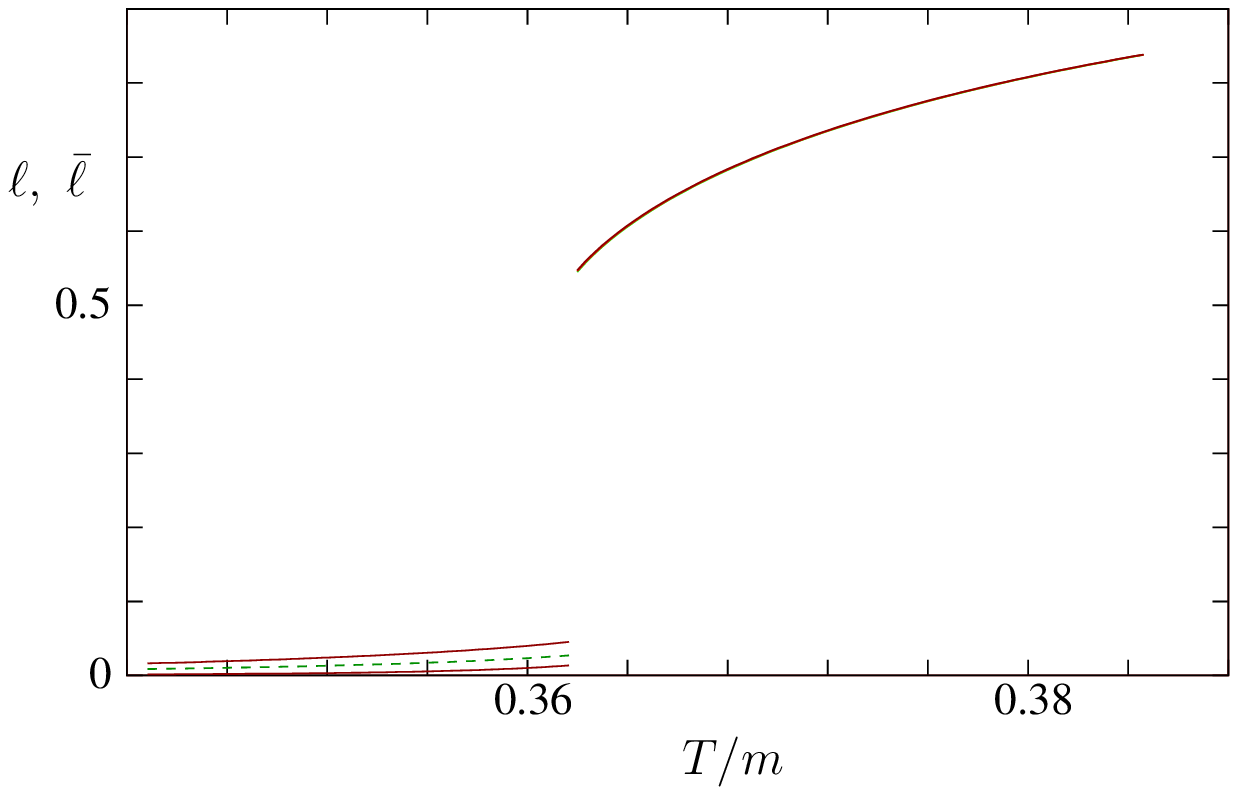}\vspace{.5cm}
\includegraphics[width=.9\linewidth]{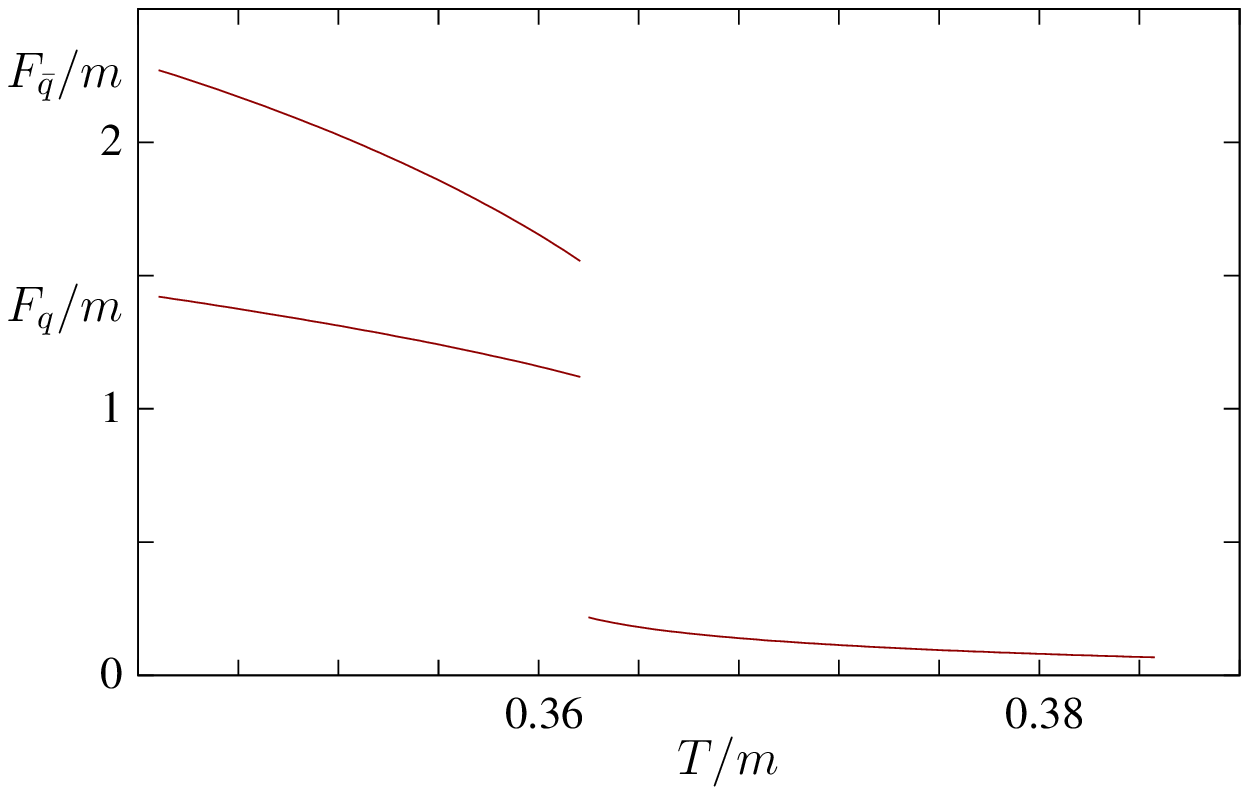}
  \caption{Upper panel: Average of the traced Polyakov loop and of its Hermitic conjugate as a function of temperature for degenerate quark masses  $M_u=M_s=4 m$ and chemical potential $\mu=0.6m$, as in \Fig{fig_saddles}. At low temperatures, the average of the Polyakov loop (higher curve) differs from that of its Hermitic conjugate (lower curve). The dotted curve corresponds to the determination of the order parameter calculated for $r_8=0$. The lower panel shows the static quark and antiquark free energies obtained from the Polyakov loops of the upper figure as $\ell=\exp(-\beta F_q)$ and $\bar\ell=\exp(-\beta F_{\bar q})$.}
\label{fig_pol_mu_real}  
\end{figure}

In our analysis, we observe that the location of the saddle point is typically at $r_8\neq0$, which induces a difference between the averages of the Hermitic conjugate Polyakov loops $\ell(\mu)$ and $\bar\ell(\mu)$ already at tree-level. As already emphasized, this is physically related to the different free energies of static quarks and antiquarks at finite $\mu$. An example is depicted in \Fig{fig_pol_mu_real}, which shows the temperature dependence of the averaged Polyakov loops in the region of first-order transition. We observe a significant difference between $\ell(\mu)$ and $\bar\ell(\mu)$ below the transition temperature, whereas they essentially agree in the high-temperature phase. In other words, the energetic price to pay for a static antiquark is much higher than that for a quark (at $\mu>0$) in the quasi-confined, low temperature phase, where the Polyakov loops are small, whereas it is essentially the same in the high-temperature deconfined phase. This can be understood analytically from an approximate calculation presented in Appendix~\ref{appsec:approx}.

We conclude this section by mentioning that previous studies of the phase diagram with background field methods in the functional renormalization group and Dyson-Schwinger approaches of Refs.~\cite{Fischer:2013eca,Fischer:2014vxa} have employed another criterion than the one used here to determine the physical properties of the system. Instead of searching for a saddle point of the function $V(r_3,ir_8,\mu)$ as we propose here, the authors of Refs.~\cite{Fischer:2013eca,Fischer:2014vxa} define the physical point as the absolute minimum of the function $V(r_3,0,\mu)$ as a function of $r_3$. We have repeated our analysis using this procedure for comparison. A clear artifact of this procedure is that on the axis $r_8=0$, the tree-level expressions of the Polyakov loops $\ell(\mu)$ and $\bar\ell(\mu)$ are equal, as already mentioned. However, we have found that both criteria give essentially the same critical temperatures in our calculation. This is illustrated in \Fig{fig_pol_mu_real}. 

That the critical temperatures are not significantly modified in these two prescriptions can be traced back to the relative smallness of the values of $r_8$ obtained by following the saddle points in our procedure; see \Fig{fig_saddles}. This, in turn, originates from the strong Boltzmann suppression of the (heavy) quark contribution to the potential, which is responsible for the departure of the saddle point from the axis $r_8=0$, as discussed in Appendix~\ref{appsec:approx}. We point out that the situation might be very different in the case of light quarks \cite{Fischer:2013eca,Fischer:2014vxa} and that different procedures for identifying the relevant extremum of the potential may have more dramatic consequences. This needs to be investigated further.

\section{Conclusions}

We have studied the phase diagram of QCD with heavy quarks at leading perturbative order in a massive extension of the Landau-DeWitt gauge in the context of background field methods. The richness of the phase diagram is reproduced for both real and imaginary chemical potential and we obtain parameter free values for dimensionless ratios of quark masses over temperature at criticality which agree well with results from lattice simulations. This extends previous studies of the confinement-deconfinement phase transition in pure Yang-Mills theories \cite{Reinosa:2014ooa,Reinosa:2014zta,Reinosa:su3} and adds to the list of lattice results that can be---at least qualitatively if not quantitatively---described by this modified (massive) perturbative scheme \cite{Tissier_10,Tissier_11,Pelaez:2013cpa,Reinosa:2013twa,Pelaez:2014mxa}. 

A natural extension of the present work is to compute the next-to-leading perturbative corrections both to the potential and to the order parameter and to study whether the results converge toward lattice values. This can be done along the lines of Refs.~\cite{Reinosa:2014zta,Reinosa:su3}. Finally, it would be of interest to investigate the phase diagram in the light quark region. However, this requires the treatment of chiral symmetry breaking and involves some refinement of the present approach. We defer such studies for future work.

\section*{Acknowledgements}

We thank N. Wschebor for collaboration on related work and for fruitful discussions.

\appendix

\section{Charge conjugation symmetry at $\mu=0$}
\label{appeq:SSB}

In the presence of a nonvanishing background field, one must be careful in assessing whether a symmetry is (explicitly or spontaneously) broken since the background field itself explicitly breaks some symmetries, e.g., the global color or the charge conjugation symmetries. To probe a possible symmetry breaking, one must consider a gauge-invariant observable, which is independent of the background field. In the present approach, any such observable ${\cal O}$ can be obtained as the value of some function ${\cal O}(r,\mu)$ at the absolute minimum\footnote{We consider the case where the physical point is the absolute minimum of the potential, relevant for the cases $\mu=0$ or $\mu\in i\mathds{R}$ discussed here.} $r_{\rm min}(\mu)$ of the potential: 
\beq
 {\cal O}(\mu)={\cal O}\big(r_{\rm min}(\mu),\mu\big).
\eeq

Let us for instance consider the case of the charge conjugation symmetry $C$, discussed in the main text. A $C$-odd observable shares the same symmetry properties as the potential $V(r,\mu)$ (such that, again, one can restrict to the elementary cell in the Cartan plane) except that\
\beq
 {\cal O}(r,\mu)=-{\cal O}(-r,-\mu).
\eeq
In particular, for $\mu=0$, we can use the color symmetry \eqn{eq_R1} to get 
\beq
 {\cal O}(r_3,r_8,0)=-{\cal O}(r_3,-r_8,0),
\eeq 
which implies that ${\cal O}(r,0)=0$ on the axis $r_8=0$ and, by symmetry, on any of the dashed axes in \Fig{fig_sym_mu=0}. Now, assuming that the charge conjugation symmetry is not broken at $\mu=0$, we have ${\cal O}(0)={\cal O}\big(r_{\rm min}(0),0\big)=0$ for any such observable. In other words, the functions ${\cal O}(r,0)$ which are odd with respect to the mirror symmetries mentioned above (i.e. $r_8\to-r_8$ in the elementary cell) must vanish at the minimum of the potential. For this to be true for all odd functions of $r_8$ (we assume that any such function can be reached from the operators of the theory), the extremum must necessary lie on the axis $r_8=0$. This shows that 
\beq
 C\,\,\,{\rm manifest}\Rightarrow r_8=0\quad\Leftrightarrow\quad r_8\neq0\Rightarrow {C}\,\,\,{\rm broken}.
\eeq 

The above conclusion can be strengthened if we make the (reasonable) assumption that every minimum of the potential $V(r,\mu)$ in the elementary cell of \Fig{fig_sym_imag} corresponds to a distinct physical state. In that case we conclude that, for any value of $\mu$, 
\beq
 C\,\,\,{\rm manifest}\Leftrightarrow r_8=0.
\eeq 
In particular this implies that the relevant extremum of the potential must lie at $r_8\neq0$ at $\mu\neq0$ since $C$ is explicitly broken.

\section{Large mass approximate formula}
\label{appsec:approx}

For the range of temperatures and quark masses studied here, we have typically $\beta M\gtrsim 6$ and $\beta m\approx 2.75$. The massive degrees of freedom are thus essentially described by classical Boltzmann statistics \cite{Kashiwa:2012wa,Lo:2014vba}. Using $\exp(-\beta\varepsilon_q)\ll1$ in the integral \eqn{eq:Weiss00}, the pure gauge potential \eqn{eq:onelooppot} can be approximated as 
\beq
 \frac{V_{\rm gauge}(r)}{T^4}\approx -\frac{{\cal F}_0(r)}{2T^4}-12f(\beta m)\ell_A(r),
\eeq
where we defined the function
\beq
 f(x)=\frac{1}{\pi^2}\int_0^\infty dy y^2e^{-\sqrt{y^2+x^2}}=\frac{x^2}{\pi^2}K_2(x),
\eeq
with $K_\nu(x)$ the modified Bessel function of the second kind and where
\beq
 \ell_A(r)=\frac{1}{8}\sum_\kappa e^{ir_\kappa}=\frac{1}{4}\left(1+\cos r_3+\cos r_++\cos r_-\right)
\eeq
is the tree-level Polyakov loop in the adjoint representation, with $r_\pm=(r_3\pm r_8\sqrt{3})/2$.

Similarly, the quark contribution \eqn{eq_potq1loop} is approximated as (we consider $N_f$ degenerate flavors with mass $M$)
\beq
 \frac{V_q(r,\mu)}{T^4}\approx -3N_ff(\beta M)\left[e^{-\beta\mu}\ell_F(r)+e^{\beta\mu}\bar\ell_F(r)\right]
\eeq
with the tree-level Pokyakov loops in the fundamental representation
\beq
 \ell_F(r)=\frac{1}{3}\sum_\rho e^{ir_\rho}=\frac{1}{3}\left(e^{-i\frac{r_8}{\sqrt{3}}}+2e^{i\frac{r_8}{2\sqrt{3}}}\cos \frac{r_3}{2}\right)
\eeq
and $\bar\ell_F(r)=\ell_F(-r)$.

In the range of temperatures and quark masses considered in this study, the Boltzmann suppression factors are $f(\beta m\approx 3)\approx 5.6\times10^{-2}$ for the massive gluon modes and $f(\beta M\gtrsim 6)\lesssim 6.2\times 10^{-3}$ for quarks. Note that in the limit of very heavy (nonrelativistic) quarks, one has $
 f(\beta M)\approx (\beta M)^{3/2}e^{-\beta M}/(\sqrt{2}\pi^{3/2})$. This gives a reasonably good estimate for the cases considered here, with $\beta M \gtrsim 6$. This relative suppression of the quark contribution as compared to that of gluons explains the fact that the critical temperatures obtained in the present work are essentially insensitive to the presence of quarks; see Table~\ref{tab:mu0}.

The relative suppression of the quark contribution also explains the smallness of $r_8$ at the extremum of the potential $V(r_3,ir_8,\mu)$ obtained in \Sec{sec:re}. Indeed, as discussed in \Sec{sec:vanishing}, the extremum of the potential always lies on the axis $r_8=0$ at $\mu=0$. A nonzero value of $r_8$ at $\mu\neq0$ is, thus, entirely due to the quark contribution and is thus controlled by the suppression factor $f(\beta M)$. 

It is an easy exercise to compute the actual value of $r_8$ as a function of $r_3$ at the extremum at leading order in $f(\beta M)$. Writing 
\beq
 V(r_3,ir_8,\mu)=V(r_3,ir_8,0)+\delta V(r_3,ir_8,\mu),
\eeq
and using the fact that $\partial_{r_8} V(r_3,ir_8,0)|_{r_8=0}=0\,\,\forall\,\, r_3$, one shows that, at leading order in $\delta V$, the value of $r_8$ at the extremum is given by
\beq
 r_8^{\rm ext.}(r_3)=-\left.\frac{\partial_{r_8} \delta V(r_3,ir_8,\mu)}{\partial^2_{r_8}V(r_3,ir_8,0)}\right|_{r_8=0}.
\eeq 
Using the approximate expressions for the potential obtained above, a simple calculation yields
\beq
\label{appeq:ppppp}
 r_8^{\rm ext.}(r_3)=\frac{16N_f}{\sqrt{3}}\frac{f(\beta M)\sinh(\beta\mu)\sin^2(r_3/4)}{\left[1-\frac{3}{\pi^2}\!\left(\frac{r_3}{2}-\pi\right)^2+18f(\beta m)\cos(r_3/2)\right]}.
\eeq
We have plotted this formula against the $r_3$ coordinate of the saddle point in \Fig{fig_saddles}, which shows that it describes well the results from the complete calculation. 
 Incidentally, \Eqn{appeq:ppppp} also explains the relative smallness of the ratio $(\ell-\bar\ell)/(\ell+\bar\ell)$ in the high-temperature phase as compared to the low-temperature phase in \Fig{fig_pol_mu_real}. This originates from the rapid increase of $r_8^{\rm ext.}(r_3)$ with increasing $r_3$ in the range of interest; see \Fig{fig_saddles}.


\begin{thebibliography}{10}



\bibitem{Borsanyi:2013bia}
  S.~Bors\'anyi, Z.~Fodor, C.~Hoelbling, S.~D.~Katz, S.~Krieg, and K.~K.~Szabo,
  Phys.\ Lett.\ B {\bf 730} (2014) 99.
  
\bibitem{deForcrand:2010ys}
  P.~de Forcrand,
  PoS LAT {\bf 2009} (2009) 010.

\bibitem{Philipsen:2010gj}
  O.~Philipsen,
  arXiv:1009.4089 [hep-lat].

\bibitem{Stephanov:2004wx}
  M.~A.~Stephanov,
  Prog.\ Theor.\ Phys.\ Suppl.\  {\bf 153} (2004) 139.
   [Int.\ J.\ Mod.\ Phys.\ A {\bf 20} (2005) 4387]

\bibitem{deForcrand:2007rq}
  P.~de Forcrand, S.~Kim, and O.~Philipsen,
  PoS LAT {\bf 2007} (2007) 178.

\bibitem{Schaefer:2007pw}
  B.~J.~Schaefer, J.~M.~Pawlowski and J.~Wambach,
  Phys.\ Rev.\ D {\bf 76} (2007) 074023.

\bibitem{Sexty:2014zya} 
  D.~Sexty,
  Nucl.\ Phys.\ A {\bf 931}, 856 (2014).
  
\bibitem{Aarts:2015kea}
  G.~Aarts,
  PoS CPOD(2014) 012.
  
\bibitem{Tanizaki:2015pua}
  Y.~Tanizaki, H.~Nishimura and K.~Kashiwa,
  Phys.\ Rev.\ D {\bf 91} (2015) 101701.

\bibitem{Fischer:2011mz}
  C.~S.~Fischer, J.~Luecker and J.~A.~Mueller,
  Phys.\ Lett.\ B {\bf 702} (2011) 438.

\bibitem{Fischer:2012vc}
  C.~S.~Fischer and J.~Luecker,
  Phys.\ Lett.\ B {\bf 718} (2013) 1036.

\bibitem{Fischer:2013eca}
  C.~S.~Fischer, L.~Fister, J.~Luecker and J.~M.~Pawlowski,
  Phys.\ Lett.\ B {\bf 732} (2014) 273.

\bibitem{Herbst:2013ail}
  T.~K.~Herbst, J.~M.~Pawlowski and B.~J.~Schaefer,
  Phys.\ Rev.\ D {\bf 88} (2013)  014007.

\bibitem{Gutierrez:2013sta}
  E.~Gutierrez, A.~Ahmad, A.~Ayala, A.~Bashir and A.~Raya,
  J.\ Phys.\ G {\bf 41} (2014) 075002.

\bibitem{Fischer:2014ata}
  C.~S.~Fischer, J.~Luecker and C.~A.~Welzbacher,
  Phys.\ Rev.\ D {\bf 90} (2014) 034022.

\bibitem{Fischer:2014vxa}
  C.~S.~Fischer, J.~Luecker and J.~M.~Pawlowski,
  Phys.\ Rev.\ D {\bf 91} (2015)  014024.

\bibitem{de Forcrand:2002ci}
  P.~de Forcrand and O.~Philipsen,
  Nucl.\ Phys.\ B {\bf 642} (2002) 290.

\bibitem{de Forcrand:2003hx}
  P.~de Forcrand and O.~Philipsen,
  Nucl.\ Phys.\ B {\bf 673} (2003) 170.

\bibitem{D'Elia:2002gd}
  M.~D'Elia and M.~P.~Lombardo,
  Phys.\ Rev.\ D {\bf 67} (2003) 014505.

\bibitem{Fromm:2011qi}
  M.~Fromm, J.~Langelage, S.~Lottini and O.~Philipsen,
  JHEP {\bf 1201} (2012) 042.

\bibitem{Pisarski:2000eq}
  R.~D.~Pisarski,
  Phys.\ Rev.\ D {\bf 62} (2000) 111501.

\bibitem{Dumitru:2003cf}
  A.~Dumitru, D.~Roder and J.~Ruppert,
  Phys.\ Rev.\ D {\bf 70} (2004) 074001.

\bibitem{Dumitru:2012fw} 
  A.~Dumitru, Y.~Guo, Y.~Hidaka, C.~P.~K.~Altes and R.~D.~Pisarski,
  Phys.\ Rev.\ D {\bf 86}  (2012) 105017.
  
\bibitem{Kashiwa:2012wa}
  K.~Kashiwa, R.~D.~Pisarski and V.~V.~Skokov,
  Phys.\ Rev.\ D {\bf 85} (2012) 114029.

\bibitem{Kashiwa:2013rm}
  K.~Kashiwa and R.~D.~Pisarski,
  Phys.\ Rev.\ D {\bf 87} (2013)  096009.

\bibitem{Nishimura:2014rxa}
  H.~Nishimura, M.~C.~Ogilvie and K.~Pangeni,
  Phys.\ Rev.\ D {\bf 90} (2014)  045039.

\bibitem{Nishimura:2014kla}
  H.~Nishimura, M.~C.~Ogilvie and K.~Pangeni,
  Phys.\ Rev.\ D {\bf 91} (2015)  054004.

\bibitem{Lo:2014vba}
  P.~M.~Lo, B.~Friman and K.~Redlich,
  Phys.\ Rev.\ D {\bf 90} (2014)  074035.

\bibitem{Reinosa:2014ooa}
  U.~Reinosa, J.~Serreau, M.~Tissier and N.~Wschebor,
  Phys.\ Lett.\ B {\bf 742} (2015) 61.

\bibitem{Reinosa:2014zta}
  U.~Reinosa, J.~Serreau, M.~Tissier and N.~Wschebor,
  Phys.\ Rev.\ D {\bf 91} (2015)  045035.

\bibitem{Reinosa:su3}
  U.~Reinosa, J.~Serreau, M.~Tissier and N.~Wschebor,
  in preparation.

   \bibitem{Tissier_10}
   M.~Tissier and N.~Wschebor,
   Phys.\ Rev.\  D {\bf 82} (2010) 101701.
  
\bibitem{Sasaki:2012bi}
  C.~Sasaki and K.~Redlich,
  Phys.\ Rev.\ D {\bf 86} (2012) 014007.
  

 \bibitem{Tissier_11}
   M.~Tissier and N.~Wschebor,
   Phys.\ Rev.\  D {\bf 84} (2011) 045018.
  
  \bibitem{Curci76}
  G.~Curci and R.~Ferrari,
  Nuovo Cimento \ A {\bf 32}  (1976) 151.

\bibitem{Pelaez:2013cpa}
  M.~Pel\'aez, M.~Tissier and N.~Wschebor,
  Phys.\ Rev.\ D {\bf 88} (2013) 125003.
  
\bibitem{Pelaez:2014mxa}
  M.~Pel\'aez, M.~Tissier and N.~Wschebor,
  Phys.\ Rev.\ D {\bf 90} (2014) 065031.

\bibitem{Reinosa:2013twa} 
  U.~Reinosa, J.~Serreau, M.~Tissier and N.~Wschebor,
  Phys.\ Rev.\ D {\bf 89} (2014) 105016.

\bibitem{Gribov77}
  V.~N.~Gribov,
  Nucl.\ Phys.\  B {\bf 139} (1978) 1.

\bibitem{Neuberger:1986vv}
  H.~Neuberger,
  Phys.\ Lett.\  B {\bf 175} (1986) 69.

\bibitem{Neuberger:1986xz}
  H.~Neuberger,
  Phys.\ Lett.\  B {\bf 183} (1987) 337.
  
\bibitem{Serreau:2012cg}
  J.~Serreau and M.~Tissier,
  Phys.\ Lett.\ B {\bf 712} (2012) 97.

\bibitem{Braun:2007bx} 
  J.~Braun, H.~Gies and J.~M.~Pawlowski,
  Phys.\ Lett.\ B {\bf 684} (2010) 262.
  
\bibitem{Serreau:2015saa} 
  J.~Serreau,
  arXiv:1504.00038 [hep-th].
  
\bibitem{Svetitsky:1985ye}
  B.~Svetitsky,
  Phys.\ Rep.\  {\bf 132} (1986) 1.

\bibitem{Roberge:1986mm}
  A.~Roberge and N.~Weiss,
  Nucl.\ Phys.\ B {\bf 275} (1986) 734.
  
\bibitem{Weiss:1980rj}
  N.~Weiss,
  Phys.\ Rev.\ D {\bf 24} (1981) 475.

\bibitem{Gross:1980br}
  D.~J.~Gross, R.~D.~Pisarski and L.~G.~Yaffe,
  Rev.\ Mod.\ Phys.\  {\bf 53} (1981) 43.

\bibitem{deForcrand:2010he}
  P.~de Forcrand and O.~Philipsen,
  Phys.\ Rev.\ Lett.\  {\bf 105} (2010) 152001.

\bibitem{Marko:2014hea} 
  G.~Mark\'o, U.~Reinosa and Zs.~Sz\'ep,
  Phys.\ Rev.\ D {\bf 90} (2014) 125021.

\end{thebibliography}
\end{document}